\documentclass{aa}  

\usepackage{graphicx}
\usepackage{txfonts}
\usepackage{xspace} 

\newcommand{\obs}[1]{\tilde{#1}}

\usepackage[dvipsnames]{xcolor}

\begin{document}

   \title{AXES-SDSS: Comparison of SDSS galaxy groups with all-sky X-ray extended sources}

  \author{S. Damsted\inst{1}
          \and
          A. Finoguenov\inst{1}
          \and
        H. Lietzen   \inst{2,3,4}
        \and
        G. A. Mamon\inst{5}
                  \and
        J. Comparat \inst{6}
        \and
        E. Tempel \inst{3,10}
        \and
        I. Dmitrieva \inst{1}
          \and
          N. Clerc\inst{7}
        \and
        C. Collins\inst{8}
         \and
        G. Gozaliasl\inst{9,1} 
                \and
        D. Eckert \inst{11}
 }

   \institute{Department of Physics, University of Helsinki,
              Gustaf Hällströmin katu 2, 00560 Helsinki, Finland\\
              \email{sanna.damsted@helsinki.fi}
         \and
         National Library of Finland, University of Helsinki, 00014 Helsinki, Finland
         \and
   Tartu Observatory University of Tartu, Observatooriumi 1, 61602 T\~oravere, Estonia
\and
Finnish Centre for Astronomy with ESO (FINCA), Vesilinnantie 5, FI-20014 University of Turku, Finland
             \and
Institut d’Astrophysique de Paris (UMR 7095: CNRS \& Sorbonne Université), F-75014 Paris, France             
             \and
             Max-Planck institute for Extraterrestrial physics, Giessebachstr, Garching 85748, Germany
         \and
    IRAP, Universit{\'e} de Toulouse, CNRS, UPS, CNES, F-31028 Toulouse, France
\and
Astrophysics Research Institute, Liverpool John Moores University, IC2, Liverpool Science Park, 146 Brownlow Hill, Liverpool L3 5RF, UK 
\and
Department of Computer Science, Aalto University, PO Box 15400, Espoo, FI-00076, Finland. 
\and
Estonian Academy of Sciences, Kohtu 6, 10130 Tallinn, Estonia
\and
Department of Astronomy, University of Geneva, Ch. d’Ecogia 16, 1290 Versoix, Switzerland
}

   \date{Received ; accepted }

 
  \abstract
   {Advances in cosmological studies require us to improve our understanding of the baryonic content of galaxy groups. The key baryonic components of groups are galaxies and hot gas, while key non-baryonic mass tracers are the velocity dispersion of galaxies and the distribution of galaxies within the group. }
   {We revisit the picture of X-ray emission of groups through the study of systematic differences in the optical properties of groups, with and without X-ray emission, and we study the effect of the large-scale density field on scaling relations.}
   {We present the identification of X-ray galaxy groups using a combination of{ ROSAT All Sky Survey (RASS)} and {Sloan Digital Sky Survey (SDSS)} data. We include a new X-ray reanalysis of RASS, covering very extended (up to a size of half a degree) sources, and we account for differences in the limiting sensitivity with respect to compact and very extended X-ray emission. We applied a screening of the identified X-ray sources, based on the optical properties, to achieve 95\% clean catalogues. We used a mock SDSS survey to understand the performance of our FoF group finder and applied the {\sc Clean} algorithm to revise group mass estimates and achieve a clean membership catalogue.}
   {X-ray groups exhibit less scatter in the scaling relations and selecting the groups based on the extended X-ray emission leads to an additional scatter reduction. Most of the scatter for the optical groups is associated with a small (6\%) fraction of outliers, primarily associated with low optical-luminosity groups found in dense regions of the cosmic web. These groups are primary candidates for the contaminants in the optical group catalogues. We find that removing only those groups from the optical group sample using optically measured properties leads to a substantial reduction in the scatter {of} the scaling relations of the optical groups. We report a dependence of both the X-ray and optical luminosity of groups {on large-scale density}, which we associate with the assembly bias. These results motivate an introduction of an additional characterization of galaxy clusters and shed light on the physical origin of anomalous clustering of galaxy clusters, found by the Dark Energy Survey (DES). 
   }
   {}


   \maketitle
%
\section{Introduction}

A new generation of experiments designed to bring our understanding of dark matter and dark energy to a new level is gaining momentum. The {\it Euclid} satellite mission has started to map the dark matter distribution in the Universe with unprecedented accuracy, using the gravitational lensing effect of its mass on the shapes of observed galaxy images.  In combination with Euclid's measurements of the Universe's expansion history using galaxy redshift measurements, these results will allow for stringent tests of the cosmological concordance model, including possible time evolution of dark energy and the validity of general relativity on cosmological scales, as well as to constrain the light neutrino masses \citep{euclid}. 

The high precision of {\it Euclid} measurements will be sensitive to detailed processes associated with galaxy formation, including the growth of black holes and their feedback (co-evolution).  \cite{Semboloni11} predicted the importance of the contribution of galaxy groups to the shear power spectrum down to  $10^{13} M_\odot$.  \cite{Debackere20} confirmed this finding through modeling the distribution of baryons in groups following observations of the stellar and hot intragroup medium. Baryonic feedback processes, either from supernovae (SNe) or active galactic nuclei (AGN), redistribute the gas, which changes the matter power spectrum, in particular, on scales where the sensitivity of cosmic shear is maximal. \cite{Semboloni11} showed that ignoring this effect leads to large biases, whereas marginalising over current model uncertainties weakens the constraining power of Euclid. The way forward is to constrain feedback directly and update its implementation in models \citep[e.g.,][]{McCarthy2017}. Groups are the best targets for resolving this issue because of their scale: on the one hand, stellar feedback processes such as galactic winds are insufficient to alter the properties of massive halos and AGN feedback becomes the dominant process through which star formation is regulated \citep[see][for a review]{eckert21}. This feedback imprints telltale signatures on the warm gas, as the group binding energy and the AGN output are similar. On the other hand, all the components of galaxy groups can be robustly determined, while groups of galaxies are the lowest mass structures that impact the shear signal on the spatial scales of interest for {\it Euclid} cosmology.

Earlier studies have suggested differences in galaxy properties based on their large-scale environment \citep{Lietzen2012, Luparello2013, Einasto2014, Poudel2016, Cohen2017, Kuutma2020}. This might influence the group's baryonic composition and one of the goals of this work is to constrain this effect. Recently, \citet{Manolopoulou2021} studied the properties of X-ray clusters in different large-scale environments and concluded that clusters of matched richness and mass tend to have higher X-ray luminosities in overdense environments.  

In this paper, we build upon this work using faint catalogues of X-ray sources, initiated by the COnstraining Dark Energy with X-ray (CODEX) survey \citep{fin20} to lower mass systems, by considering the spectroscopic group catalogues available from Sloan Digital Sky Survey (SDSS). The results of the study presented in this paper are limited to the $0.015<z<0.2$ range. At lower redshifts ($z<0.015$), the apparent size of groups exceeds a degree and the concept of source detection changes, as the largest angular scales for detecting the group X-ray emission considered in this work only cover the core of the groups; thus, for the sake of simplicity we leave this to future work. {The upper limit on the redshift is imposed by the SDSS optical group catalogue, due to the loss of survey sensitivity with respect to a group detection, with further insights on this presented in Sect.\ref{mock}.}
We also make a comparison of the optical groups to a subsample with X-ray emission.  This deliberately leaves out a class of X-ray luminous, optically hard-to-identify groups, known as large magnitude gap groups. They are easier to study with surveys having a large range of optical magnitudes covered, such as CFHTLS and COSMOS \citep{Gozaliasl14, Gozaliasl2019}. Therefore, we have prioritized the available spectroscopic characterization of the optical groups instead, as they can be used to remove any doubt that those are not real systems, in addition to the large sample size of the SDSS group catalogue.

We assume a {\it Planck} cosmology with the Hubble constant $H_0=67.8$~km\,s$^{-1}$~Mpc$^{-1}$ (also noted as h=0.678 in the paper), matter density, $\Omega_m=0.308$, and dark energy density, $\Omega_\Lambda=0.692$ \citep{Planck2016}. We use a natural-based log ($ln$) to perform the analysis and tabulating the results of the fits while using $\log_{10}$ for plotting. {The evolution of scaling relations is governed by the critical density, which is scaled as $E_z=(\Omega_m (1+z)^3+\Omega_\Lambda)^{0.5}$.}

\section{Data}\label{data}

\subsection{Optical spectroscopy}

We used the optical friends-of-friends (FoF) group catalogue by \citet{Tempel2017},  based on the SDSS Data Release 12 (DR12; \citet{Blanton2005,Adelman-McCarthy2008,Padmanabhan2008,Eisenstein2011, Alam2015}). The catalogue covers the largest contiguous area of the main galaxy sample of the SDSS Legacy Survey. \citet{Tempel2017} constructed the group catalogue using the FoF algorithm based on the spectroscopic flux-limited galaxy catalogue. The galaxies were limited to Petrosian $r$-band magnitude brighter than 17.77. Their redshifts were corrected for the motion with respect to the cosmic microwave background and limited to $z<0.2$. The full catalogue contains 584\,449 galaxies and 88\,662 groups with at least two members. 

The galaxy group characterization is strongly affected by interlopers \citep[e.g.,][]{Mamon2013}, and to address this issue, we have introduced an additional step applying the membership cleaning code {\sc Clean} of \citet{Mamon2013}, retaining the galaxies within $R_{200c}$, which results in a final catalogue of 5\,801 groups with five or more members and a corresponding clean membership catalogue of 57\,352 galaxies.

\begin{figure*}[t] 
\centering
\begin{tabular}{c c}
\resizebox{0.45\textwidth}{!}{\includegraphics{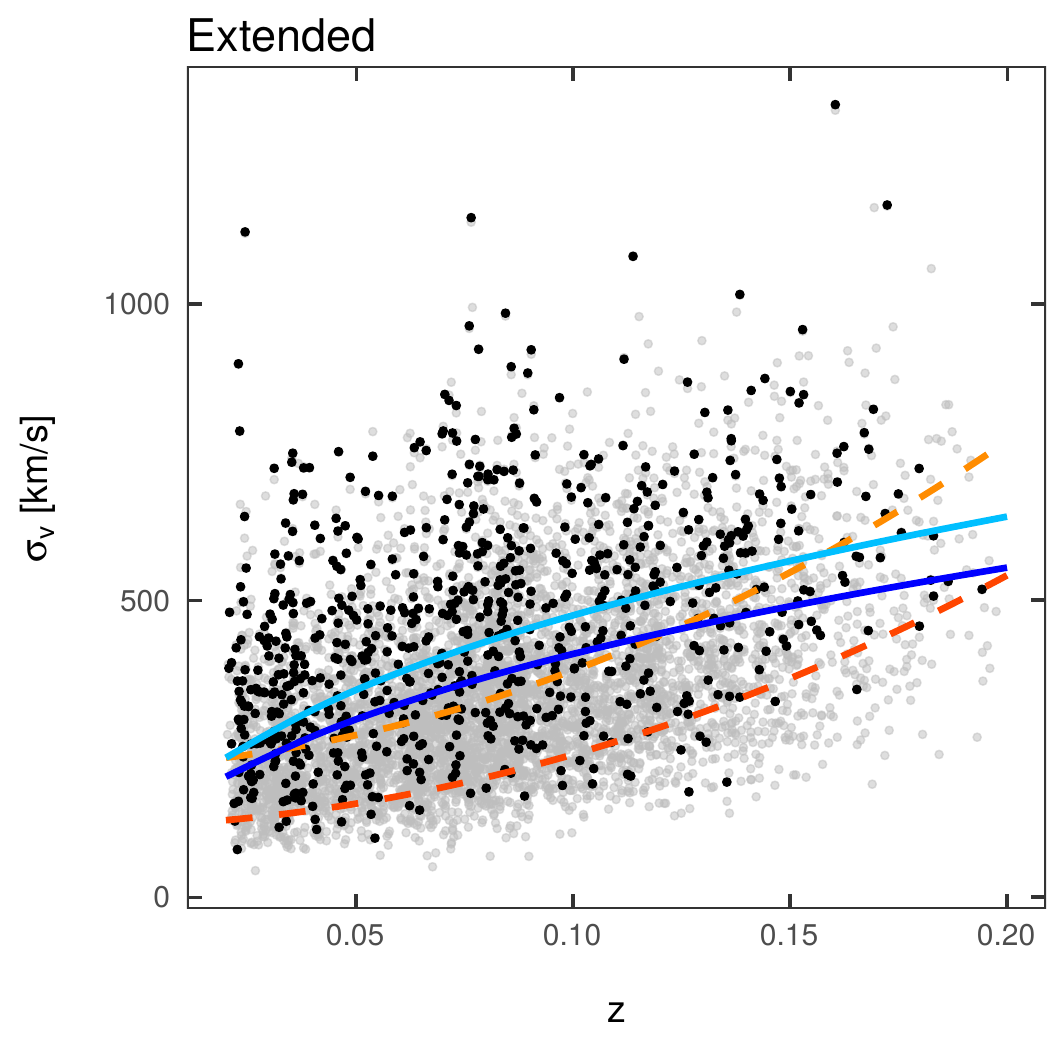}}&
\resizebox{0.45\textwidth}{!}{\includegraphics{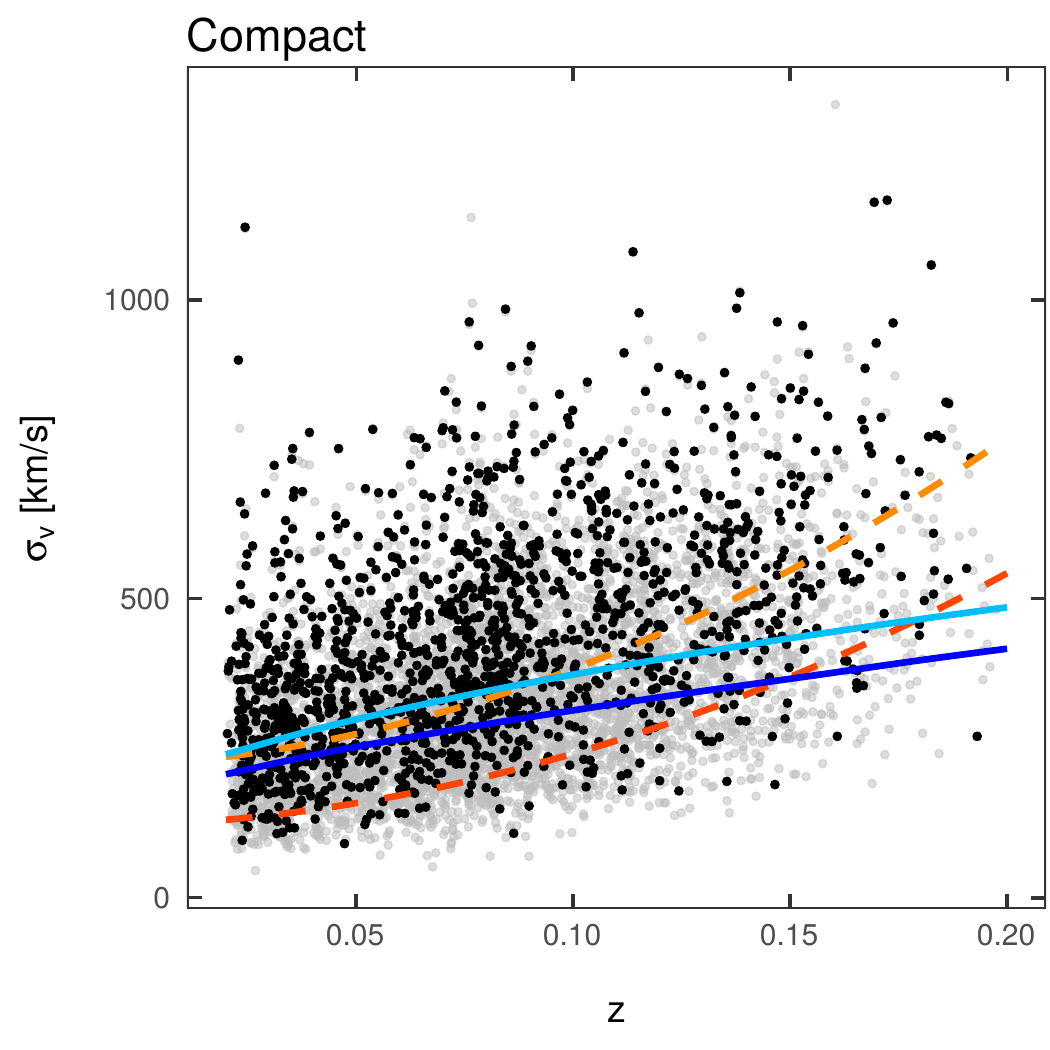}}
\end{tabular}
\caption{Sample of extended (left panel) and compact (right panel) X-ray groups (black points). All optical groups are shown with grey points in both panels. Completeness limits for the FoF groups are shown in red (10\%) and orange (50\% completeness) dashed lines, and for the X-ray groups in dark blue (10\%) and light blue (50\%) solid lines. Systems below the 10\% X-ray completeness limit are removed from the catalogues. We applied the same extended source cleaning to all catalogues used in the analysis.}
\label{completeness}
\end{figure*}

Optical groups can be characterized by several properties, such as those defined in \citet{Tempel2014}. In this paper, we use their number of members as a data quality tracer, $r$-band luminosity redefined to a uniform limit of $M_{\rm r}<-19.5$, and velocity dispersion, which we have refined using the clean membership. We use the gapper method \citep{Beers90} to estimate the velocity dispersion, suitable for the data in hand. In addition, we report velocity dispersions computed using mean absolute deviation \citep[MAD;][]{Beers90} in the catalogue. Since using five-member galaxies to estimate the velocity dispersion is uncertain, for the analysis presented in this paper, we use a sub-catalogue of 1\,512 groups with ten or more members.

\citet{Tempel2017} constructed the luminosity weighted galaxy density field as described in \citet{Liivamagi2012}. The density field was smoothed with four smoothing scales: 1.5~Mpc, 3~Mpc, 6~Mpc, and 10~Mpc. The densities are expressed in units of the mean density, which is $\overline{D}=0.01227\times 10^{10}M_\odot \mbox{Mpc}^{-3}$. In this paper, we would like to explore the link between the large-scale density of galaxies and the mass of the group we are studying. These densities trace the large-scale clustering of matter, which in the linear approximation of the large-scale structure growth, scales with the product of the bias of the groups with the bias of galaxies used to compute the density field. We take the density estimate coming from the largest scale, 10 Mpc, to reduce the contribution of the halo itself and the concentration of galaxies next to it.  The bias of the groups depends on group mass as well on the epoch of group formation. So, while serving as an independent non-baryonic mass tracer, the density field has additional sensitivity, compared to the velocity dispersion.
Properties of the groups as a function of the large-scale density have been studied by \cite{meinasto24}.

There is a growing variety of optical group catalogues, based on SDSS data, which pursue improvements in the sensitivity to the low halo mass of the group as well as the robustness of the mass estimate \citep[e.g.,][]{maggie, Tinker2021, galweight}. The sensitivity of our X-ray data does not benefit from these improvements, nevertheless, we checked that our assessment of group mass would not change if we were to use the MAGGIE catalogue \citep{maggie} for the systems in common. For this work, we rely on the group membership catalogues,  which are best understood for FoF groups \citep{Tempel2017}, the identification of merging systems {and the large-scale density field estimates}.
In this work, we have adopted a non-baryonic cleaning of the X-ray counterparts using group velocity dispersion. The cleaning of the catalogue is an essential step to ensuring the purity of the identification \citep{klein19}, while the use of non-baryonic mass proxies is an attempt to remove the baryonic properties of halos from the selection. The velocity dispersion has been remeasured for each group in the catalogue, using the optical centre, as it is available for all the groups, and applying the {\sc Clean} algorithm \citep{Mamon2013} to remove the outliers and provide the velocity dispersion estimate within the aperture of $R_{200c}$. 

\subsection{X-rays}

The X-ray work presented in the paper is an extension of the CODEX clusters survey \citep{fin20} toward lower redshifts and masses. The construction of the compact X-ray source catalogue is identical to CODEX, which is based on applying the advanced wavelet filtering techniques used on X-ray images. The denomination "compact" refers to wavelet scales sampling RASS PSF, with the largest wavelet scale that is included being $6^\prime$. Compared to the CODEX catalogue, the only change is in the identification method, which in this work consists of matching the source list to the SDSS spectroscopic group catalogue \citep{Tempel2017}. At the same time, CODEX is based on applying the redMaPPer at the position of X-ray sources. 
Following the approach of \cite{Xu18}, in identifying large sources as low-z galaxy groups, and confirmed by the dedicated consideration of galaxy group emission in \citet{Kaefer2019}, we  included large X-ray sources; these were found using wavelet scales of 12 or 24 arcminutes in the identification work. The X-ray source catalogue forms the All-sky X-ray Extended Sources (AXES) project. The motivation for considering large angular scales is to create a source catalogue sensitive to the baryonic content of groups at the outskirts. We do not introduce further source filtering based on, for instance, the source extent, as this imposes unnecessary model assumptions on the shape of the X-ray emission, which could bias our comparison to the optical catalogues. 
Consistently with this approach, we quote the full X-ray flux of the source. In both compact and extended X-ray source catalogues, the flux is estimated using background subtracted sky images and wavelet reconstruction only defines the region of flux extraction. In computing the extrapolated flux, we account for the aperture effects and report the flux extraction radius and a ratio between the measured and extrapolated flux. We refer to the representative follow-up of extended sources performed by the XMM-Newton project XGAP \citep{Eckert24}. 
The centre of X-ray emission is determined by computing the barycentre of the wavelet reconstruction by running {\sc SExtractor} \citep{bertin} separately on compact and extended image reconstructions, suppressing the signal-to-noise filtering in {\sc SExtractor}, but allowing for the deblending and rejection of sources occupying less than 5 pixels, which are $45^{\prime\prime}$ on a side. Compact X-ray emission reconstructions include the wavelet scales of 1.5, 3, and 6 arcminutes, and extended emission -- 12 and 24 arcminutes.  The detection of X-ray emission on different scales has a physical origin. Given a typical cluster core size of $0.1R_{200c}$ and typical extent of detectable X-ray emission of $R_{500c}\sim 0.65R_{200c}$, and given the redshift range of our survey, cluster cores extend to 2--10 arcminutes and cluster X-ray emission extends to half a degree. This separation of dominant contribution to the X-ray emission on core vs outskirts is at the root of the difference in the catalogue properties. 

We needed to select the scoring parameter to perform the X-ray source identification. The velocity dispersion provides a non-baryonic ranking, while the distance between optical and X-ray centres has to be limited to avoid chance identification. Given our goal to compare the groups with and without X-ray emission, we restricted our identification to high-velocity dispersion groups, by applying a cut based on the 10\% completeness of X-ray catalogues as well as avoiding $z<0.015$. For the compact sources, we use the same richness cut applied to the CODEX cluster catalogue \citep[][{Eq. 1}]{fin20} and expressed it as a cut in the velocity dispersion using the calibrations of \cite[][{Table 2 and Fig. 11}]{Damsted23}: 
\begin{equation} \label{eqn:cut1}
P^{\rm RASS}(I | \obs{\sigma_v}, z)=\theta (\obs{\sigma_v}-375(z/0.15)^{0.38}),
\end{equation}
where $\theta$ denotes a step-function and $\obs{\sigma_v}$ is observed velocity dispersion in km s$^{-1}$. Here, P is the introduced probability of selection (I) given $\obs{\sigma_v}$. The corresponding sensitivity limit for extended sources, based on our analysis of the X-ray source detection and flux extrapolation, is:
\begin{equation} \label{eqn:cut2}
P^\mathrm{RASS-ext}(I_{ext} | \obs{\sigma_v}, z)=\theta (\obs{\sigma_v} -  490(z/0.15)^{0.44}),
\end{equation}
and it is less sensitive than the selection of compact sources at all redshifts considered here.

Estimating the velocity dispersion of groups with few galaxies is uncertain \citep{saro13}. To reduce the possible effect of this uncertainty, our analysis of scaling relations only includes groups with ten or more galaxies. 
A natural choice for a matching distance should encompass a typical offset between the optical and X-ray centres for well-studied cases. The separation of 200 kpc is well justified by the results on gas sloshing, as studied at X-rays with Chandra and XMM-Newton \citep{HFB, AM06, Johnson10,Zuhone13,V14,Ge20}. 
Further refinement in the distance could include a rescaling based on velocity dispersion, but it would introduce a covariance with observational errors on velocity dispersion. Given the limited span in the sampled system masses, we consider using a constant limit on the maximum offset acceptable. Our selection agrees well with the offset distribution found for eROSITA sources \citep{seppi23}. 
The total number of compact sources identified is 1121, where we removed systems having multiple counterparts and kept the closest X-ray source to the centre of the optical group.

The completeness limits are shown in Fig. ~\ref{completeness}.
The X-ray properties of the groups are included in the online catalogues, as described in Table \ref{tab:catalogue_columns}. A large number of sources is identified using both compact and extended scales of emission. There is still a large difference in the used apertures of X-ray flux estimates, $R_{\rm E}$ (which is the same as the extent of X-ray emission). Extended sources require almost no extrapolation and their flux estimates are most robust. The amount of extrapolation increases with decreasing redshift, resulting in differences in redshift scaling between Eqs.~\ref{eqn:cut1} and \ref{eqn:cut2}. 
Comparison to the redMaPPer cluster identification of CODEX \citep{fin20}, using spectroscopic galaxy groups allows us to assess smaller mass systems, having  $\sigma_v<400$ km s$^{-1}$. The X-ray sensitivity of RASS data toward the detection of groups as compact sources extends to $z<0.18$ and as extended sources to the $z<0.1$ range. The lower limit on the redshift of the survey (0.015) is set by a large difference in the optical group size and our X-ray detection scales, which would otherwise lead to additional differences between the optical and X-ray groups, complicating the interpretation of the results.  At the lowest redshift boundary of the survey, we reach the required sensitivity for the detection of groups with dispersion down to 156 km s$^{-1}$ for compact and 178 km s$^{-1}$ for extended X-ray sources. While cleaning the extended X-ray source catalogue might appear excessive, on scales of half a degree, there is at least a 4\% contribution of galactic sources, such as supernova remnants to the source catalogue, {which is revealed by an excess of AXES sources in the galactic plane \citep[][]{axes2mrs}}. The removal of a chance association with abundant low-mass groups is of a larger importance, which has been addressed by a cut using the velocity dispersion. High redshift cluster contamination is addressed using a comparison to the CODEX primary sample released in \cite{fin20}. For further details on the purity and completeness of the identification, we refer the reader to Kosowski et al. (in prep.). 
In Fig.~\ref{completeness} we also show the 50\% completeness curves, useful for understanding the regime where completeness corrections are not dominant. The figure illustrates that the rate of recovering completeness is better at X-rays compared to optical group finders, as previously discussed in \citet{Finoguenov09}. This is due to a much steeper dependence of X-ray luminosity on mass, compared to the galaxy number density. Also, an increase in the total number of galaxies with survey depths depends on the group galaxy luminosity function. In contrast, the expected number of X-ray photons for a given source increases proportionally to X-ray sensitivity.

In the case of redMaPPer identification with {the Spectroscopic Identification of eRosita Sources} (SPIDERS) follow-up, the completeness level of 375~km~s$^{-1}$ is achieved at $z=0.1$, superseding the completeness level of the spectroscopic group finder. Thus, the SPIDERS cluster catalogue will offer additional sources in the same area. The advantages of spectroscopic group finders consist of their purity at lower halo masses, which is the focus of our study here, and group searches at $z<0.1$, a redshift range characterized by strong projection effects in red sequence cluster finders. For example, the authors of the DES paper even considered limiting redMaPPer studies to $z>0.2$ \citep{Abbott2020}. Thus, we are justified to present the results in the full redshift range of the SDSS group catalogue.

Taking the extent of the X-ray emission as an estimate of the largest scale of the emission used to determine the centre and rationing it by the significance of the flux estimate, the uncertainty of X-ray positions for 90\% of sources are within $0.15 R_{200c}$ and $0.4 R_{200c}$ for compact and extended sources, respectively. This position uncertainty scales with redshift: for the compact X-ray sources, it is within 0.1 and 0.2 $R_{200c}$ for redshifts below and above 0.06 for 90\% of sources and for extended sources the corresponding positional uncertainty is within 0.3 and 0.5 $R_{200c}$. 
Finally, we have eyeballed all the extended sources to flag source confusion and contamination cases. The contamination is determined by finding an enhancement in the X-ray emission at the position of an unassociated SDSS galaxy group or redMaPPer cluster. We have either changed the flux extraction area to avoid contamination (and assigned a flag equal to 2 to the source) or removed the source from the analysis in cases of strong confusion and where the significance of flux estimate from the uncontaminated area is low (below 95\% confidence level). We assigned flag 3 to the sources where the contamination could not be fully excluded but was also not obvious. Flag 1 sources have no contamination. All the X-ray flux measurements are corrected by comparing the extraction area to $R_{500c}$ and following the procedures outlined in \cite{finoguenov07}. We removed from the analysis the compact sources that require extreme (more than a factor of 12) X-ray flux corrections. In most cases, no corrections are made for extended sources.

\subsection{Mock SDSS catalogue}\label{mock}

In assessing the sensitivity for the spectroscopic group search, we created an SDSS mock based on SMDPL and MDPL2 Multidark simulations \citep{Klypin2016}.
The halos and subhalos were detected and catalogued using \textsc{rockstar} \citep{Behroozi2013b}.
Stellar mass and star formation rates are predicted for halos and subhalos with the \textsc{UniverseMachine} model \citep{Behroozi2019}.
We projected all-sky light cones as in \citet{Comparat2020} to obtain angular coordinates and redshifts.
We note that in groups and clusters, there is a slight underdensity of high-mass satellites in dark matter halos as well as an underdensity of sub-halos toward the centre \citep{Comparat2017}.
Based on the COSMOS catalogue \citep{Laigle2016}, we calibrated the relation between stellar mass, star formation, and the observed absolute magnitude in $r$-band.
Finally, using the SDSS magnitude limit, we drew a spectroscopic SDSS mock catalogue from this mock one.

We ran the improved version of FoF \citep{Tempel2017} on the mock and applied the {\sc Clean} algorithm to clean membership and to obtain velocity dispersion estimates. We used the gapper estimates in this paper. To understand the completeness and purity of the catalogue, we matched the FoF group catalogue to the halo catalogue of simulations, allowing for redshift displacements on the order of $0.003(1+z)$, mis-centering up to $0.5 R_{200c}$ radii and selecting the most matching mass in case several counterparts are found. 
The computation of sensitivity is done by comparing the cumulative fraction of detected vs the total number of halos as a function of halo mass in the mock catalogue. We transferred this to the velocity dispersion space using the scaling relations of \cite{Carlberg1997}. Our results on the scatter of the measured velocity dispersion versus halo mass compare well with those previously reported by \citet{Old2015} and we do not repeat them here.
We estimated the fraction of outliers as those outside the $3\sigma$ range of the log-normal distribution of scatter in derived vs simulated halo properties. The five-member catalogue has 15\% impurity and the ten-member catalogue has a 10\% impurity, with a much higher measured velocity dispersion with respect to the simulated halo mass. A stacking analysis reveals that this is an effect of contamination from a two-halo term. We noted that the cleaning of the catalogue becomes much more efficient and reaches 99\% once we are able to assess the true velocity dispersion, as contaminants have values much below the boundary of this study of 300~km\,s$^{-1}$. Thus, we conclude that the role of X-ray selection is to access the true properties, which therefore serves as an efficient cleaning of the catalogue. 
The mock catalogue allows us to compute the completeness limits of the FoF group finder. We computed the 10\% and 50\% completeness limits by finding the halo mass at which 10\% and 50\% of the halos have been correctly detected. 
The resulting completeness limits can be approximated with polynomials $11266z^2+486z+221$ (in km s$^{-1}$) for the 50\% completeness limit and $9079z^2+294z+121$ for the 10\% limit. We show the resulting 10\% and 50\% completeness curves in Fig. ~\ref{completeness}. The results of our study correspond to the area of over 50\% completeness of the FoF group finder. Our calculation of the FoF catalogue sensitivities allows us to assess the limitations of this study.

\section{Results}\label{results}

Our catalogue of X-ray groups introduces additional properties to the optical group catalogues and allows us to look for a difference in the optical properties of groups depending on their X-ray selection. The common characteristics of the groups we can access are group luminosity, $L_{\rm r}$, velocity dispersion, $\sigma_{\rm v}$, and the location within the large-scale structure, which we have traced with the luminosity density measured with the 10~Mpc smoothing scale, named $D_{10}$ in \cite{Tempel2017}. In studying the scaling relations, we selected the following order of parameters against which the study is done: $\sigma_v$, $L_X$, and $L_r$. Once a property has been selected as a base, it is normalized by a round value close to the median, to remove the degeneracy between the normalization and the slope of the scaling relation. Here, $L_X$ has been normalized by $10^{43}$ erg s$^{-1}$, $L_r$ by $35\times10^{10}L_\odot$, and $\sigma_v$ by 400 km s$^{-1}$. {On the Y-axis, $L_r$ is in units of $10^{10} L_\odot$.} To account for the evolution of the scaling relations, we use $L_X E^{-1}_z$, $L_r E_z$. To study the scaling relations, we followed the Bayesian regression routine, \textit{linmix}, {from} \citet{Kelly2007} using N=3 Gaussian components. Three regression parameter equations were used for calculating line fits in the \textit{linmix} routine:

\begin{equation}\label{eq:linmix eta}
    \eta_i=\alpha+\beta \xi_i +\epsilon
,\end{equation}
where $\alpha$ is the intercept, $\beta$ is the slope, and $\epsilon$ is intrinsic random scatter about the regression. Here, $\epsilon$ is assumed to be normally distributed with a zero mean and variance $\sigma^2$ ($N(0,\sigma)$). Thus, we have:
\begin{equation}\label{eq:linmix x}
    x_i=\xi_i + \epsilon_{x,i} ,
\end{equation}
where $\xi_i$ are the data points with $x_{\epsilon_{x,i}}$ errors:
\begin{equation}\label{eq:linmix y}
    y_i=\eta_i+ \epsilon_{y,i},
\end{equation}
where $\epsilon_{y,i}$ is the error in $\eta_i$ (both are data).
%
%

We started with a comparison of properties of optical and X-ray groups, using {the large-scale structure} density-optical luminosity and optical luminosity-velocity dispersion, as well as the {large-scale structure} density-velocity dispersion scaling relations. We computed the errors for large-scale structure density, optical luminosity and velocity dispersion, as $D_{10, err}=\sigma(D_{gal})$, $L_{r,err}=L_{r,gal,min}$, and $\sigma_{v,err}=0.91*\sigma_v / \sqrt{N_{gal}-1}$. The environment was defined for each galaxy and reported as $D_{gal}$, so the error term for the group is the variance of environments for its galaxies. The error on the total luminosity is set to trace the depths of the galaxy catalogues included. In estimating the errors, it is important to differentiate between the well and poorly-defined systems, while any general uncertainty associated with a property is described through the scatter term of the scaling relation. For example, the {\it linmix} estimate of the scatter of the optical luminosity includes additional scatter due to the completeness of the SDSS survey and efforts on completing the spectroscopic sampling may potentially be rewarded by a smaller value of scatter measured. 

As our SDSS group galaxy sample is flux-limited, we limited the calculation to a subsample of galaxies with $M_{\rm r}<-19.5$ to provide a meaningful comparison of group optical luminosities. The choice of the threshold is to include low-mass groups, while still having enough survey volume where this calculation is complete. Using only $M_{\rm r}<-19.5$ galaxies offers a uniform definition of $L_{\rm r}$ without survey-specific redshift trend in the data. Following \citet{Zheng21}, we also require a minimum of five galaxies that fill the $M_{\rm r}<-19.5$ condition. An alternative way of introducing a redshift-dependent extrapolation of optical luminosity requires an understanding of how the extrapolation procedure influences the scatter.  Hereinafter, for the computation of optical luminosity we will use the volume-limited subsample of galaxies. In contrast, we use all (magnitude-limited sample) galaxies in the computation of the velocity dispersion.
The adopted calculation of the optical luminosity limits the study to $z<0.06$, which we selected for the main analysis. Given the strong evolution of the scatter of $L_X$ below $z<0.15$, found by \citet{Damsted23}, we offer a comparison to  the $0.06<z<0.15$ scaling relations in {Appendix \ref{Appendix A}}.

\begin{figure}[ht]
    \centering
    \includegraphics[width=\hsize]{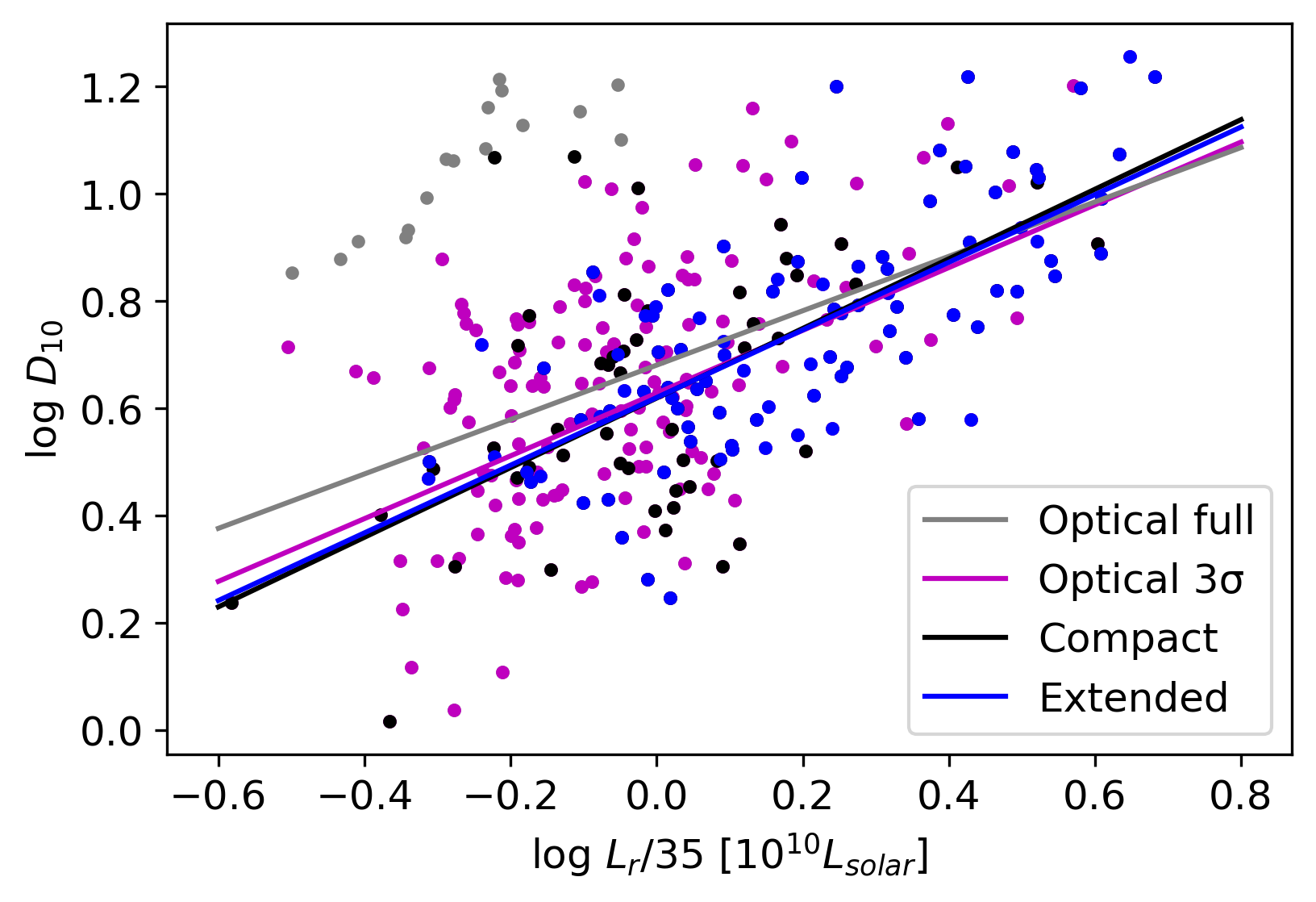}
    \caption{Large-scale structure density versus the optical luminosity data points and \textit{linmix} scaling relations. Grey and magenta show the optical sample without and with a $3\sigma$ cut. Black and blue show the samples for compact and extended X-ray sources.}
    \label{fig: scatter D-Lr full}
\end{figure}

\begin{figure}[ht]
    \centering
    \includegraphics[width=\hsize]{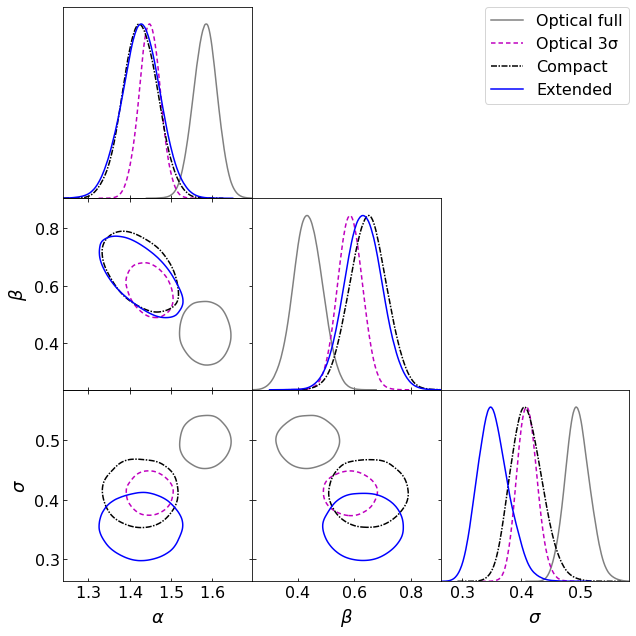}
    \caption{Large-scale structure density vs. optical luminosity regression parameters for the $0.015<z<0.06$ groups. Here, $\alpha$ stands for intercept, $\beta$ -- for slope and $\sigma$ -- for the intrinsic scatter. The 2D contours show the 90\% significance levels. An optical sample without a 3$\sigma$ cut is marked with solid grey and with the cut with dashed magenta lines. The compact X-ray sample is marked with dash-dotted black and extended with solid blue. The X-ray clusters exhibit a steeper relation and lower scatter.}
    \label{fig: triangle D-Lr full}
\end{figure}

\begin{figure}[ht]
    \centering
    \includegraphics[width=\hsize]{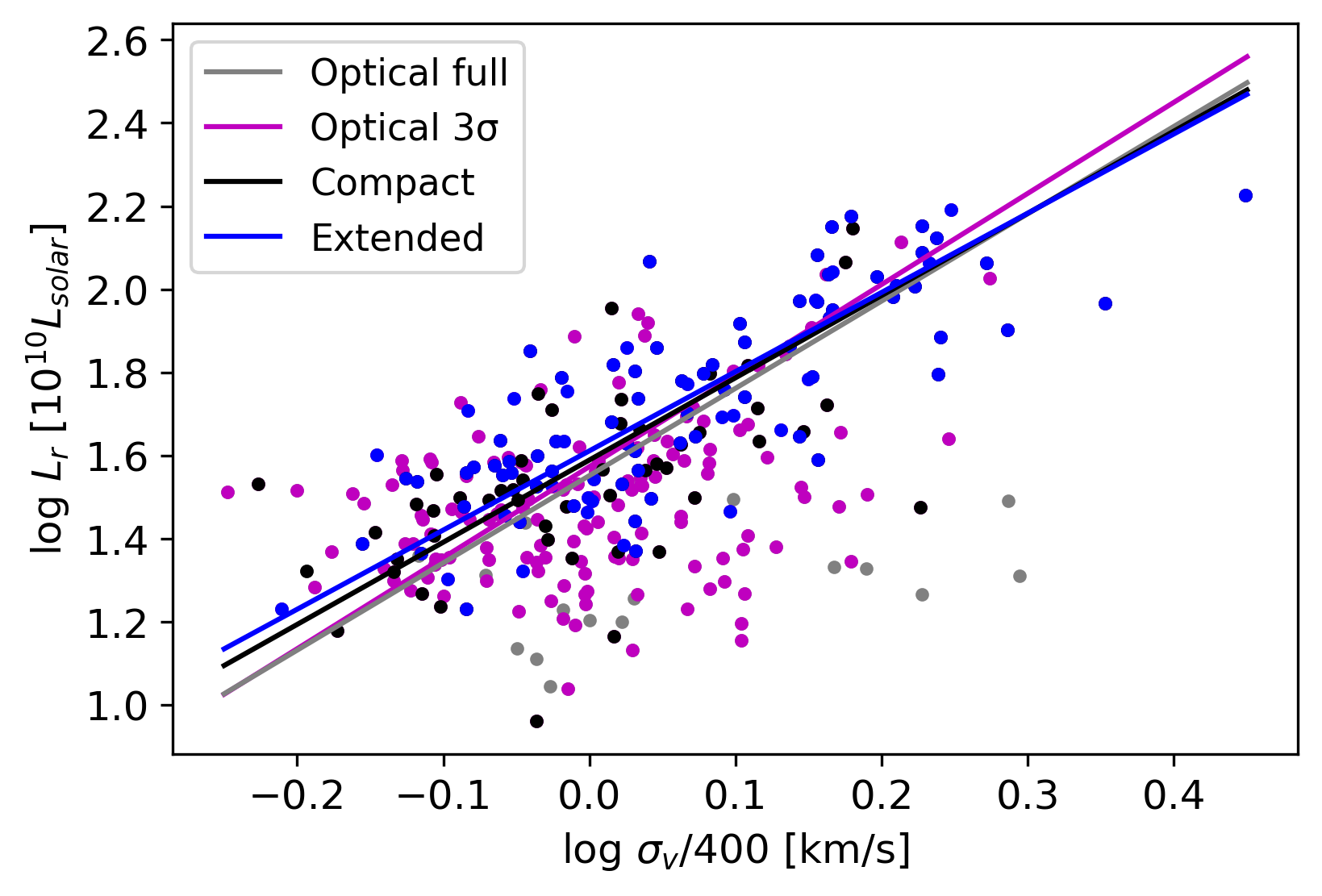}
    \caption{Optical luminosity vs. velocity dispersion. Details are the same as in Fig. \ref{fig: scatter D-Lr full}.}
    \label{fig: scatter Lr-vdisp full}
\end{figure}

\begin{figure}[ht]
    \centering
    \includegraphics[width=\hsize]{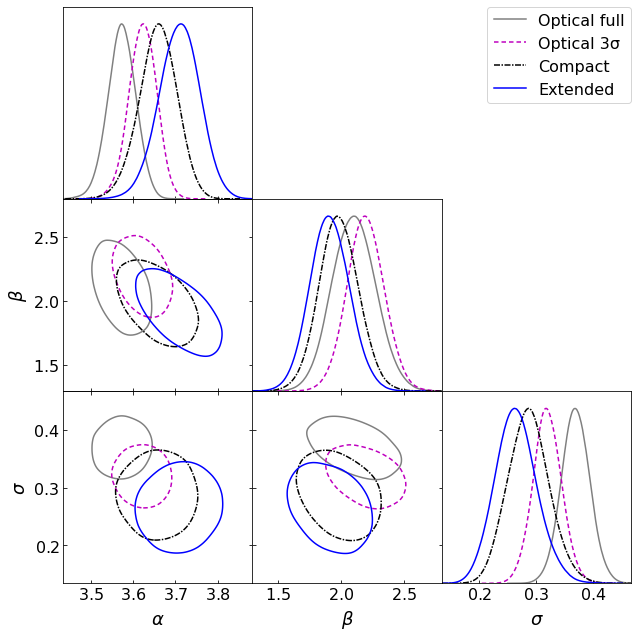}
    \caption{Optical luminosity vs. velocity dispersion. Details are the same as in Fig. \ref{fig: triangle D-Lr full}}
    \label{fig: triangle Lr-vdisp full}
\end{figure}

\begin{figure}[ht]
    \centering
    \includegraphics[width=\hsize]{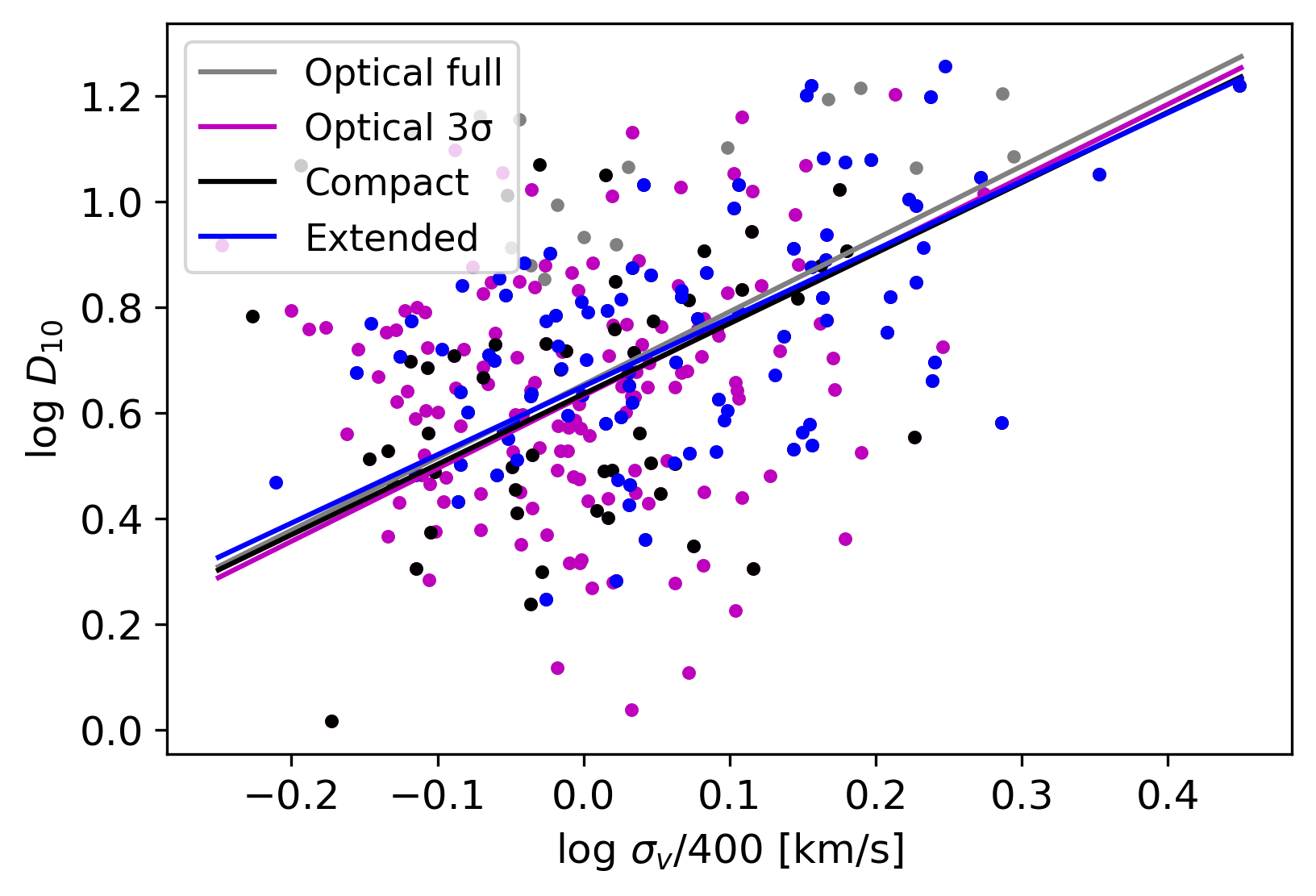}
    \caption{Large scale structure density vs. velocity dispersion. Details are the same as in Fig. \ref{fig: scatter D-Lr full}.}
    \label{fig: scatter D-vdisp Lr cut}
\end{figure}

\begin{figure}[ht]
    \centering
    \includegraphics[width=\hsize]{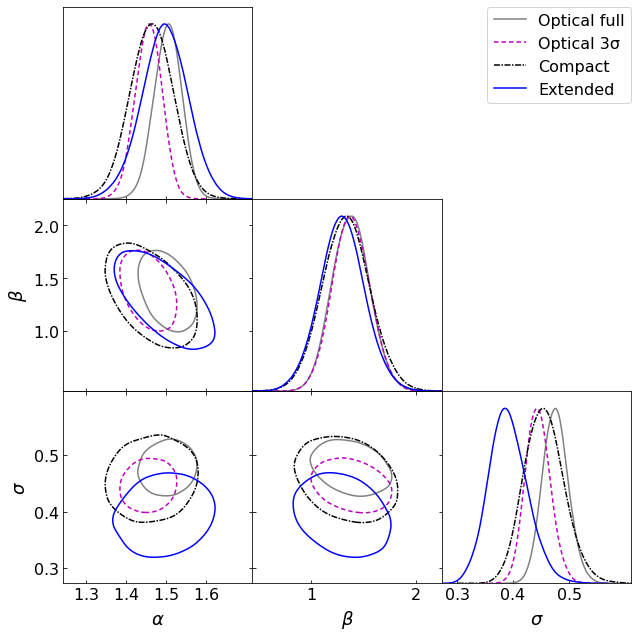}
    \caption{Large scale structure density vs. velocity dispersion. Details are the same as in Fig. \ref{fig: triangle D-Lr full}}
    \label{fig: triangle D-vdisp Lr cut}
\end{figure}

The selection of X-ray groups based on the compact X-ray emission has been the common way of selecting galaxy groups and alternative samples were not large enough. With the introduction of our extended source catalogue, we offer the first attempt to see whether X-ray detection affects the conclusion on the optical properties of the groups. Most (80\%) groups with extended X-ray emission also contain the compact X-ray source, but not the other way around; this is because the compact X-ray catalogues contain contributions from relaxed low-mass groups \citep{Eckert11} or groups hosting an AGN. We present the results for both samples to see the improvements introduced by the extended X-ray source selection.  
In comparing to the optical groups, we tested two hypotheses in looking at the trends given by the data: one is that the non-X-ray detected groups contain fake groups and the other is that non-X-ray detected groups have a higher fraction of baryons locked in stars. We expect that fake optical groups contain a lower optical luminosity and are located in denser environments, while groups with a higher fraction of baryons in stars would have a higher optical luminosity for a given velocity dispersion. Figure \ref{fig: scatter D-Lr full} demonstrates that non-X-ray groups that contribute to the extra scatter are located at very high densities. Figure \ref{fig: scatter Lr-vdisp full} shows that non-X-ray groups that strongly deviate from the scaling relation have lower optical luminosity. These two observations lead us to conclude that a difference in the scatter in the optical properties of groups is driven by the fake groups. 
We use a Z-score method for outlier removal in the optical group catalogue using $D-L_r$ scaling relation and performing a $3\sigma$ cut, using the parameters of the scaling relation for extended X-ray sources. {Overall, 18 outlier groups were removed,  leading to a 6\% reduction in the number of optical groups.} 
The culled optical sample has much better similarity to the X-ray sample, though deviations toward lower optical luminosity are still present there, as made evident in the $L_r-\sigma_v$ scaling relation {(Figs. \ref{fig: scatter Lr-vdisp full} and \ref{fig: triangle Lr-vdisp full})}. The observed larger scatter in the $L_r-\sigma_v$ relation for the optical groups, compared to the X-ray groups, is in agreement with (and extends to lower X-ray luminosities) the study of \citet{popesso04}, whose sample of 114 clusters (with 90\% of sources having $L_\mathrm{0.1-2.4\; keV}>5\times10^{43}$ erg s$^{-1}$ was based on the RASS point source catalogue of \cite{Voges99}, which is close to our compact source list. 
The slope of the  $L_r-\sigma_v$ scaling relation is flatter than 3 for any group sample {(see Table \ref{table:linmix values 1})}. This implies a decrease in the M/L ratio on the group scale in this sample, consistent with a general trend of higher star formation efficiency in groups compared to clusters \citep{leauthaud12}. A visual impression for a steepening of the relation at velocity dispersions above 550 km s$^{-1}$ is consistent with the original finding of \cite{girardi02}. While the full optical catalogue exhibits a large scatter around this trend, the X-ray catalogues do not reveal a need for additional effects. Nevertheless, there are plenty of well-behaved optical groups without X-ray emission. Optical luminosity is the best mass proxy in terms of the scatter, compared to a full $L_X$, as evident from the scatter measured with respect to $\sigma_v$ {(Figs. \ref{fig: scatter Lr-vdisp full}, \ref{fig: triangle Lr-vdisp full}, \ref{fig: scatter LX-vdisp Lr cut}, and \ref{fig: triangle LX-vdisp Lr cut})} or large-scale density {(Figs. \ref{fig: scatter D-Lr full}, \ref{fig: triangle D-Lr full}, \ref{fig: scatter D-LX Lr cut}, \ref{fig: triangle D-LX Lr cut}, and Table \ref{table:linmix values 1})}. As discussed in \cite{Damsted23}, the X-ray luminosity becomes a better mass tracer at $z>0.15$. Instead, one advantage of the X-ray selection is in the removal of contamination in optical group catalogues, which allows us to efficiently recover the properties of genuine groups. 
Analysis of the $D_{10}-\sigma_v$ relation in Figs. \ref{fig: scatter D-vdisp Lr cut}-\ref{fig: triangle D-vdisp Lr cut} shows that X-ray groups occupy the same parameter space as the optical groups, with extended X-ray sources exhibiting marginally lower scatter. We interpret this result as indicating that selection effects can be reduced when matching the samples based on the velocity dispersion. 

\begin{figure}[ht]
    \centering
    \includegraphics[width=\hsize]{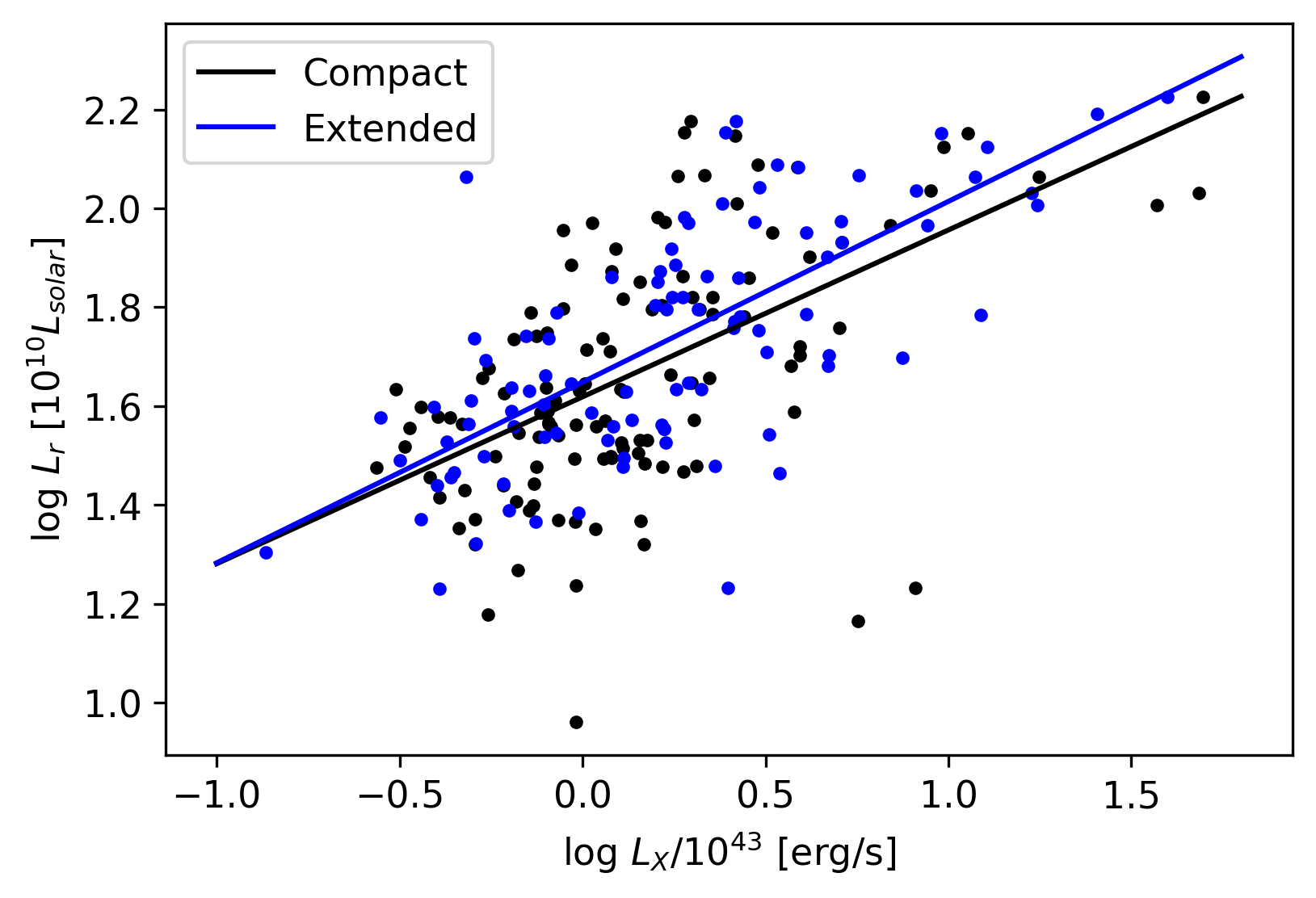}
    \caption{Optical luminosity vs. X-ray luminosity. Black shows the compact and blue the extended X-ray source samples.}
    \label{fig: scatter Lr-LX}
\end{figure}

\begin{figure}[ht]
    \centering
    \includegraphics[width=\hsize]{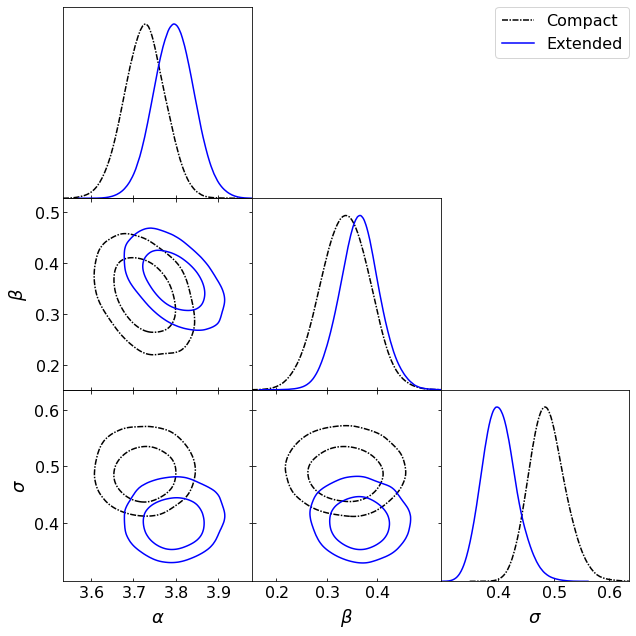}
    \caption{Optical luminosity vs. X-ray luminosity. Black dash-dotted lines show the compact X-ray source sample and blue solid lines the extended one. The 2D contours show the 68\% and 95\% significance levels. The other details are the same as in Fig. \ref{fig: triangle D-Lr full} }
    \label{fig: triangle Lr-LX full}
\end{figure}

\begin{figure}[ht]
    \centering
    \includegraphics[width=\hsize]{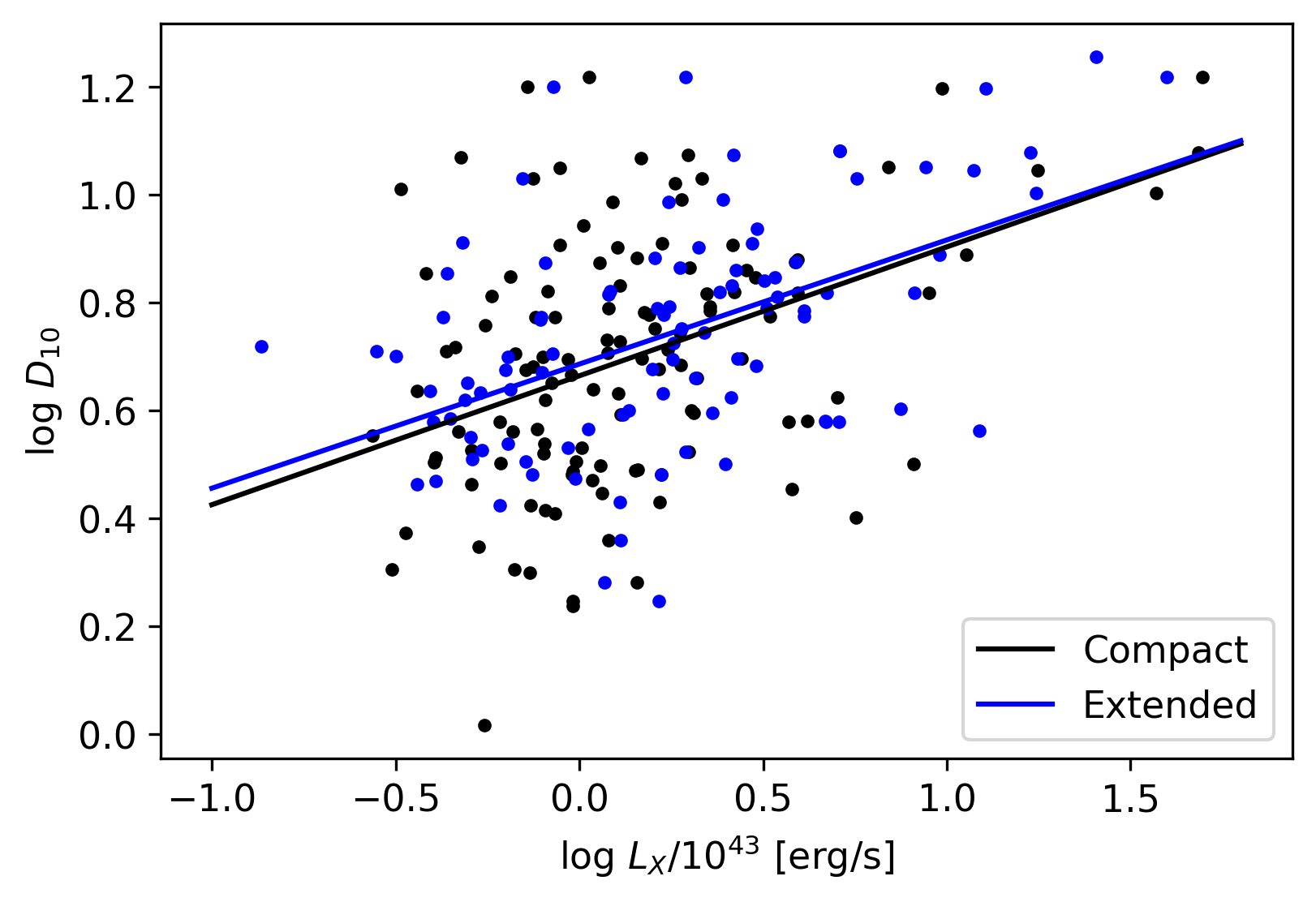}
    \caption{Large-scale structure density vs. X-ray luminosity. The  compact (black) and  extended X-ray (blue) source samples are shown.}
    \label{fig: scatter D-LX Lr cut}
\end{figure}

\begin{figure}[ht]
    \centering
    \includegraphics[width=\hsize]{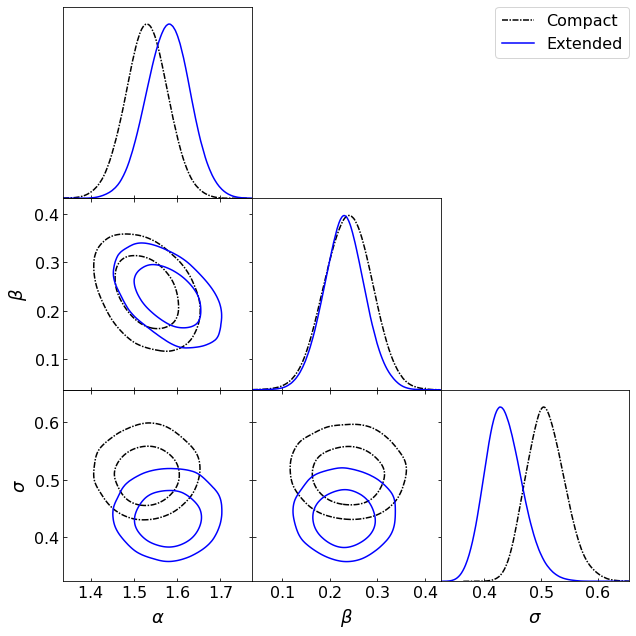}
    \caption{Large-scale structure density vs. X-ray luminosity. Details are the same as in Fig. \ref{fig: triangle Lr-LX full}.}
    \label{fig: triangle D-LX Lr cut}
\end{figure}

\begin{figure}[ht]
    \centering
    \includegraphics[width=\hsize]{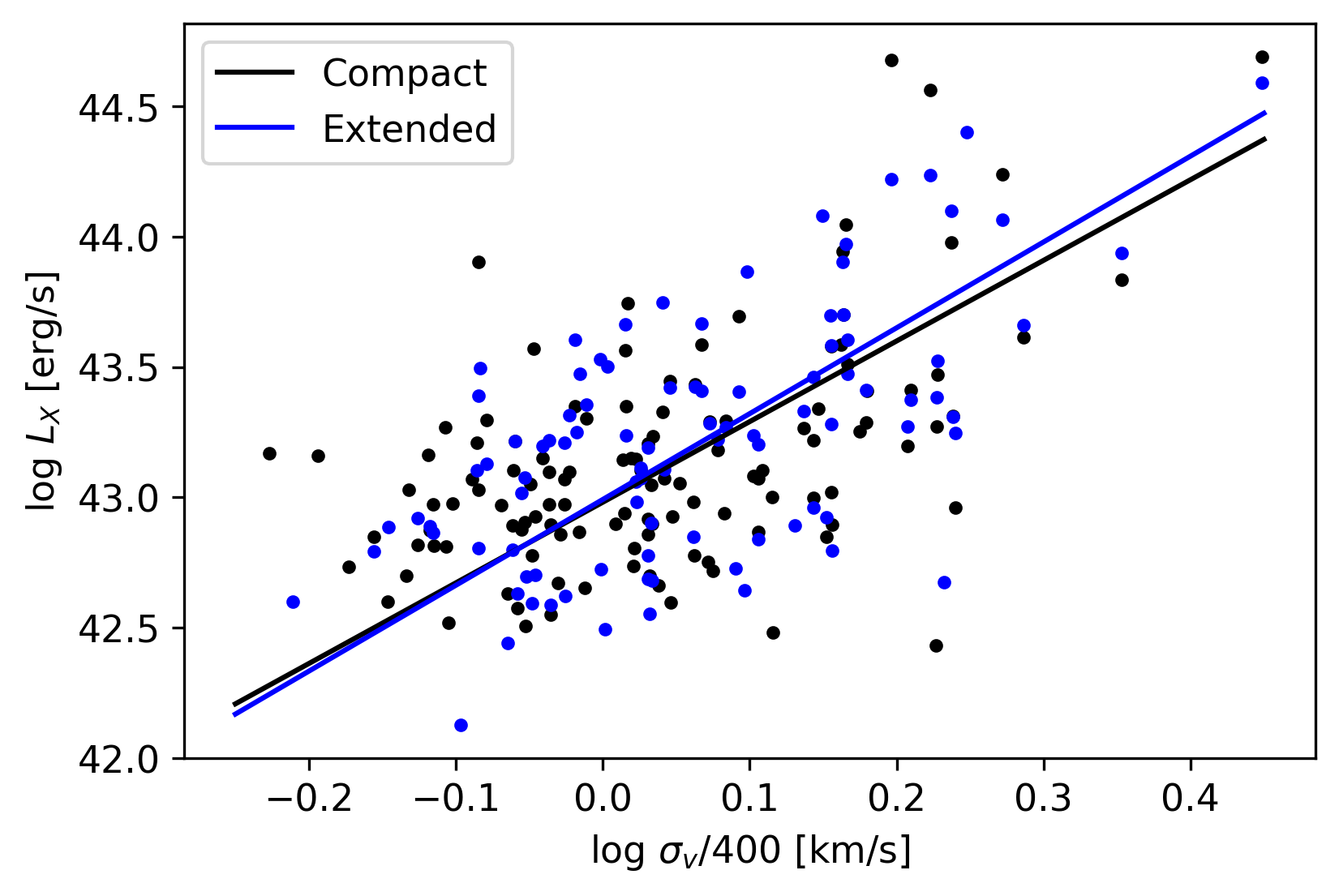}
    \caption{X-ray luminosity vs. velocity dispersion. The  compact
(black) and  extended X-ray (blue) source samples are shown.}
    \label{fig: scatter LX-vdisp Lr cut}
\end{figure}

\begin{figure}[ht]
    \centering
    \includegraphics[width=\hsize]{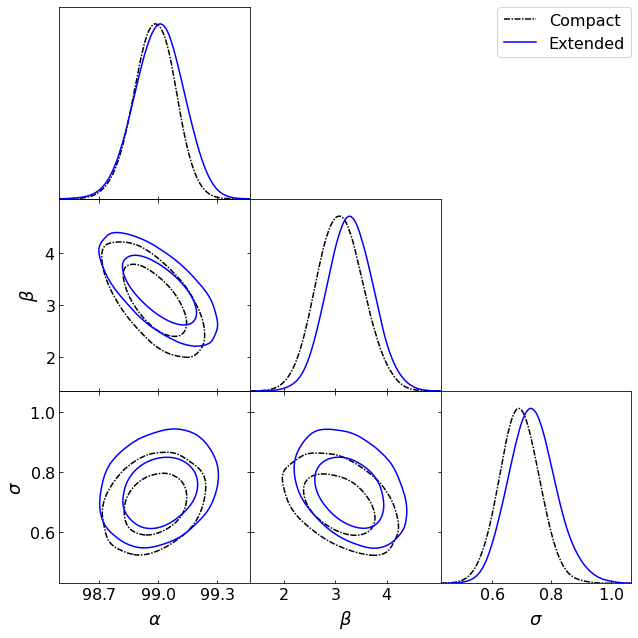}
    \caption{X-ray luminosity vs. velocity dispersion.  Details are the same as in Fig. \ref{fig: triangle Lr-LX full}.}
    \label{fig: triangle LX-vdisp Lr cut}
\end{figure}

The X-ray properties of compact and extended samples are compared in Figs. \ref{fig: scatter Lr-LX}--\ref{fig: triangle Lr-LX full}. The slopes and normalization of the scaling relations are in agreement and the main difference consists of the extended source exhibiting a lower scatter. Nevertheless, the level of scatter for extended sources is higher than what is found for $z>0.15$ X-ray clusters in \cite{Damsted23}, confirming their results on the strong increase in scatter in full $L_X$ toward low redshifts. 
The $L_{X}-\sigma_v$ scaling relation provides insights into the mass calibration of the sample. 

\begin{figure}[ht]
    \centering
    \includegraphics[width=\hsize]{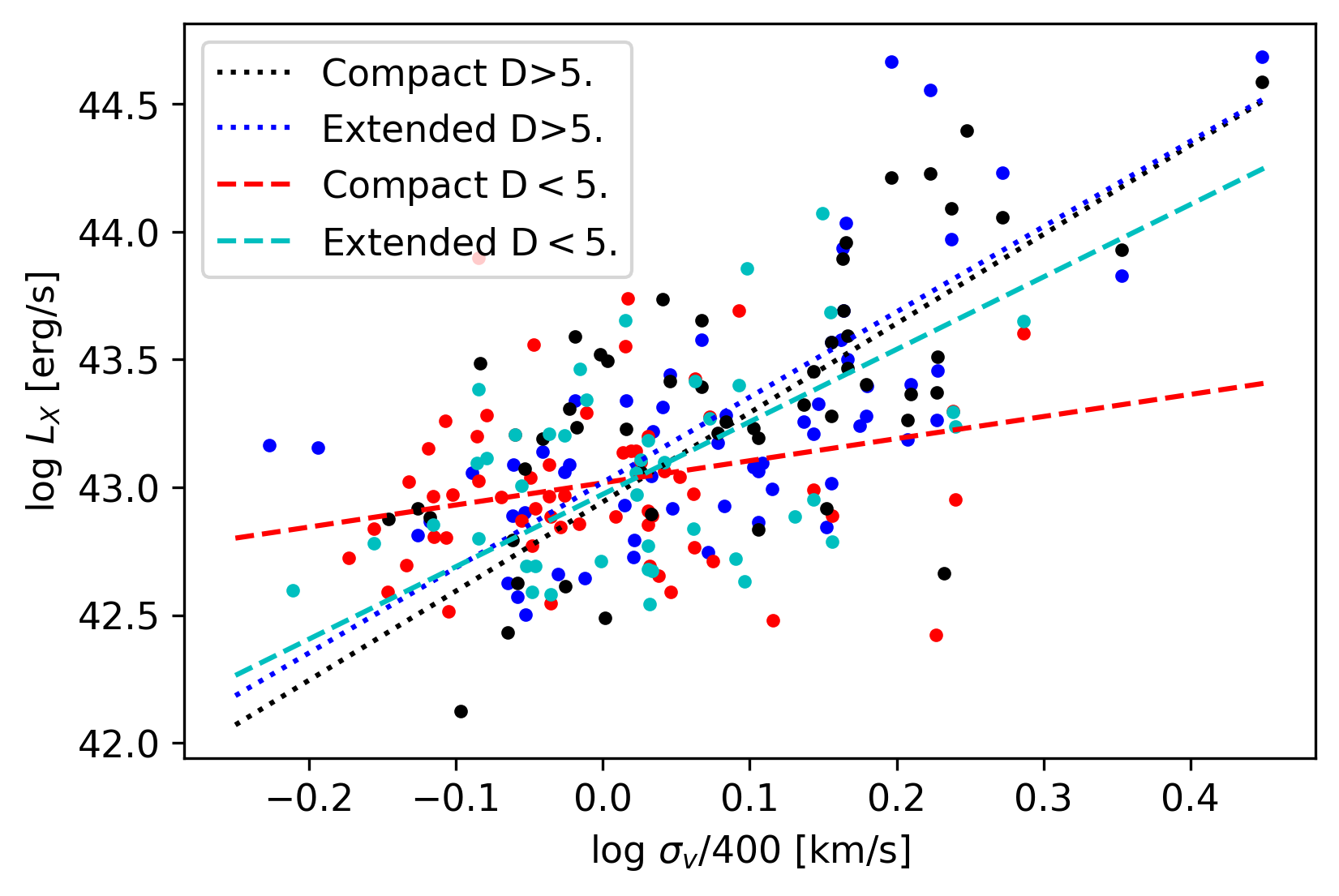}
    \caption{X-ray luminosity vs. velocity dispersion with a division in large scale structure density at $D_{10}=5.$ Black circles show the compact and blue circles show the extended X-ray source samples, with $D_{10}>5.$ with their scaling relations as dotted lines. Red and cyan circles show respectively the compact and extended samples having $D_{10} \leq 5.$ with scaling relations as dashed lines.}
    \label{fig: scatter LX-vdisp D cut}
\end{figure}

\begin{figure}[ht]
    \centering
    \includegraphics[width=\hsize]{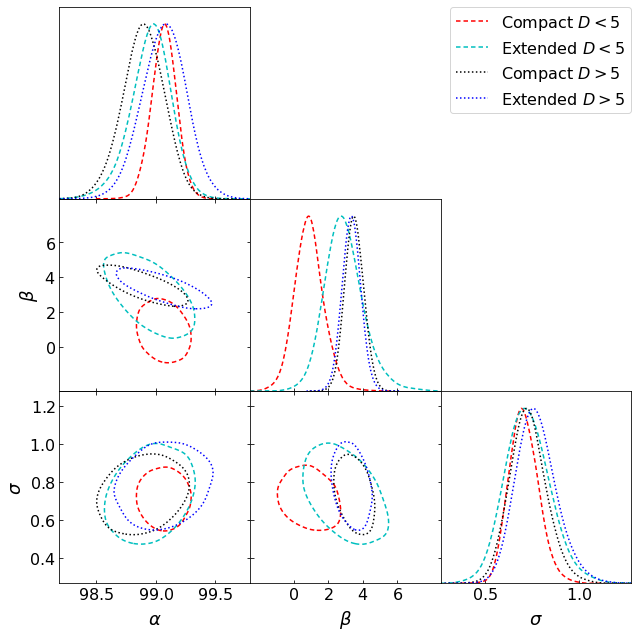}
    \caption{X-ray luminosity vs. velocity dispersion with large scale structure density cut at $D_{10}=5.$ Black circles show the compact and blue circles show the extended X-ray source samples, with $D_{10}>5.$ with their scaling relations as dotted lines. Red and cyan circles show respectively the compact and extended samples having $D_{10} \leq 5.$ with scaling relations as dashed lines. Other details are the same as in Fig. \ref{fig: triangle D-Lr full}.}
    \label{fig: triangle LX-vdisp D cut}
\end{figure}

\begin{figure}[ht]
    \centering
    \includegraphics[width=\hsize]{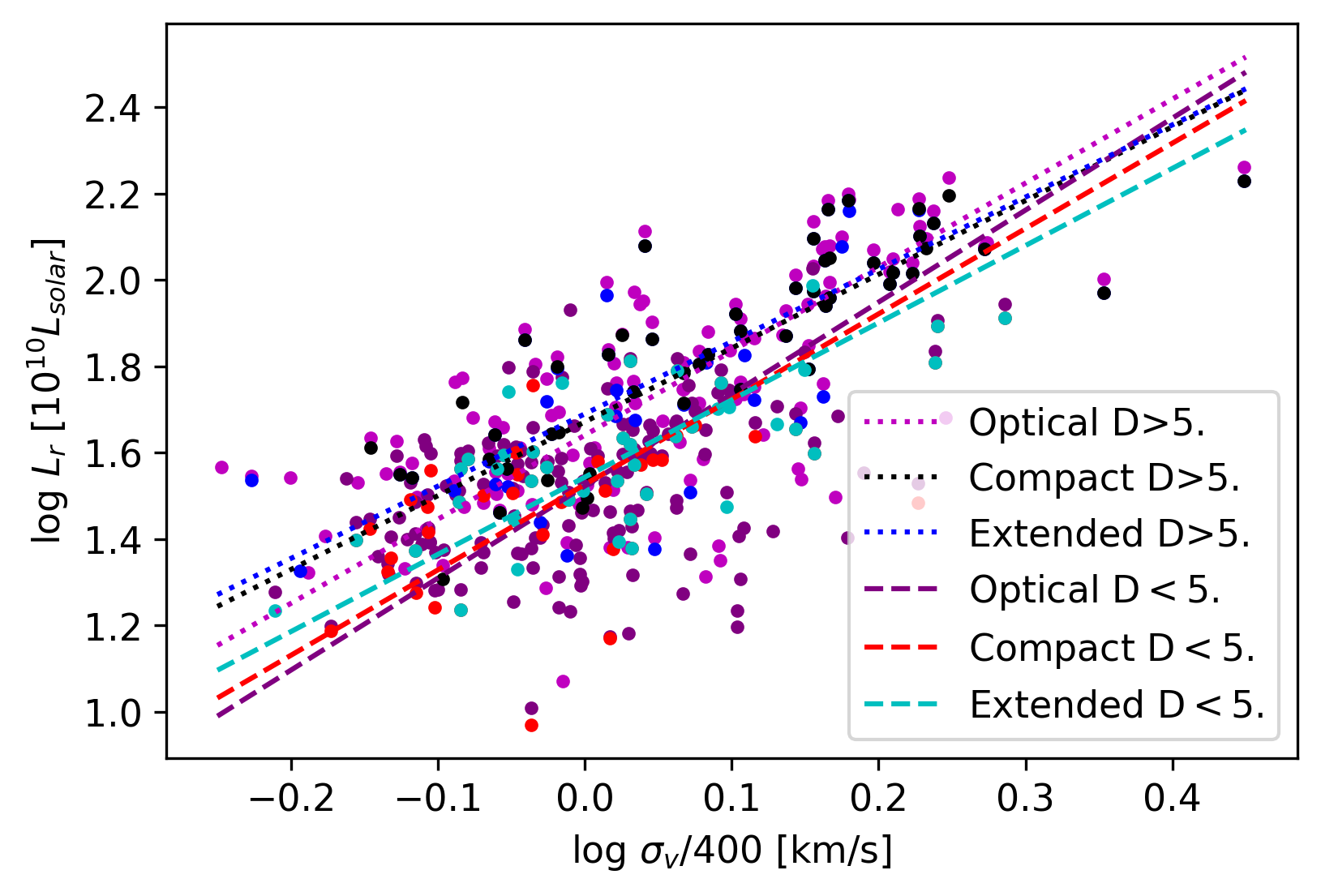}
    \caption{Optical luminosity vs. velocity dispersion with a division in large scale structure density at $D_{10}=5.$ Optical sample is marked with a magenta line: dashed for the lower and dotted for the higher large-scale structure density samples. Other details are the same as in Fig. \ref{fig: scatter LX-vdisp D cut}.}
    \label{fig: scatter Lr-vdisp D cut}
\end{figure}

\begin{figure}[ht]
    \centering
    \includegraphics[width=\hsize]{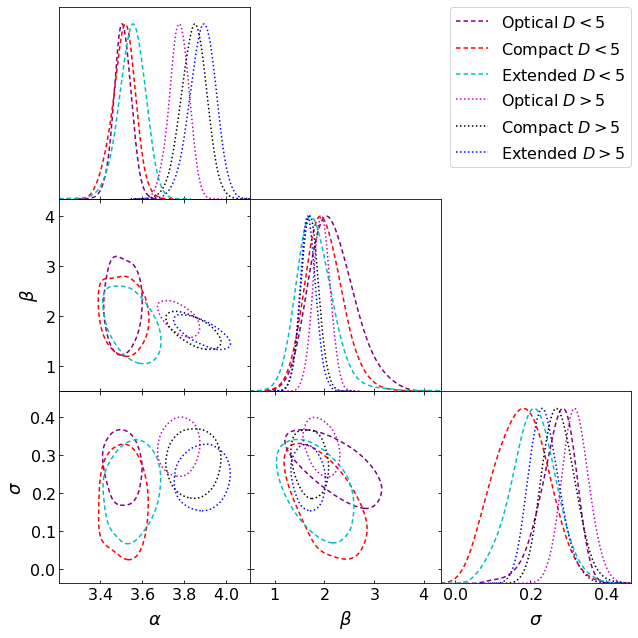}
    \caption{Optical luminosity vs. velocity dispersion with large scale structure density cut at $D_{10}=5.$ Optical sample is marked with a magenta line: dashed for the lower and dotted for the higher large-scale structure density samples. Other details are the same as in Fig. \ref{fig: triangle LX-vdisp D cut}.}
    \label{fig: triangle Lr-vdisp D cut}
\end{figure}

\begin{figure}[ht]
    \centering
    \includegraphics[width=\hsize]{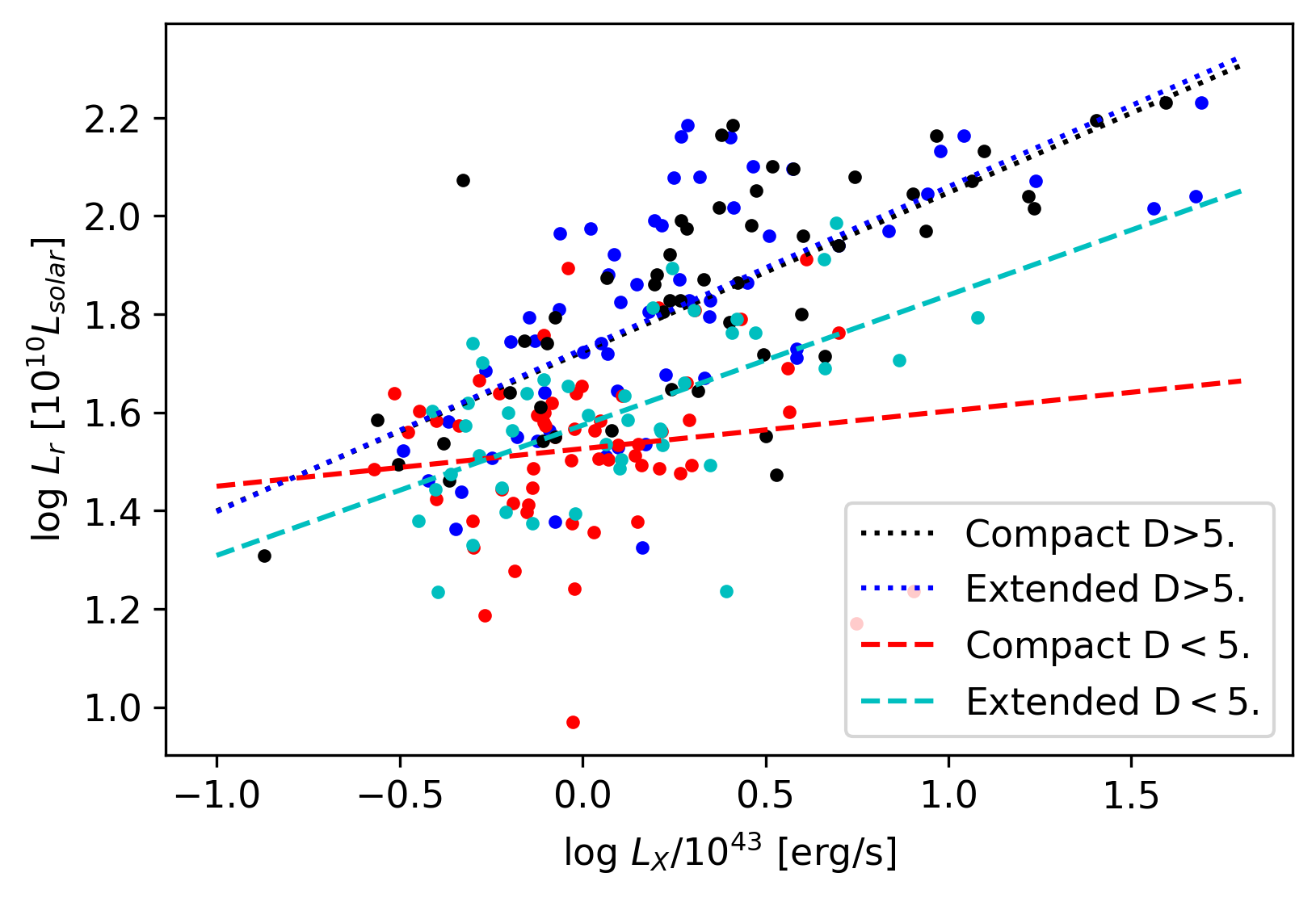}
    \caption{Optical luminosity vs. X-ray luminosity with a division in large scale structure density at $D_{10}=5.$ Details are the same as in Fig. \ref{fig: scatter LX-vdisp D cut}.}
    \label{fig: scatter Lr-LX D cut}
\end{figure}

\begin{figure}[ht]
    \centering
    \includegraphics[width=\hsize]{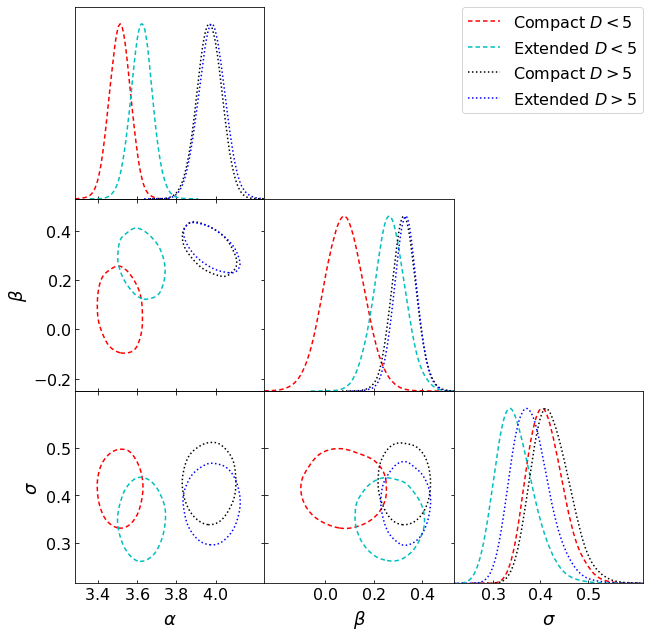}
    \caption{Optical luminosity vs. X-ray luminosity with large scale structure density cut at $D_{10}=5.$ Details are the same as in Fig. \ref{fig: triangle LX-vdisp D cut}.}
    \label{fig: triangle Lr-LX D cut}
\end{figure}

In Figs. \ref{fig: scatter LX-vdisp D cut}-\ref{fig: triangle Lr-LX D cut}, we study the scaling relations split by the large-scale density. We find a sensitivity of both X-ray and optical scaling relations to the large-scale structure density. Groups found in dense environments exhibit a higher X-ray luminosity and higher optical luminosity for a given velocity dispersion.{ We note that for the extended X-ray sources, the uncertainty on parameter estimates is too large to either support or reject this conclusion.} Given that our catalogues are spectroscopic and the presence of X-ray emission confirms the groups, the projection effects are minimal. As both X-ray and optical luminosity increase with increasing density, density-dependent effects introduce a positive covariance of the scatter between optical luminosity and X-ray luminosity, compensating for the negative covariance between these two properties expected from the baryonic conversion effects. This finding supports the statement of \cite{puddu22} done for similarly massive systems and similar redshifts that there is no anti-covariance of X-ray and optical properties \citep[observed in massive $0.15<z<0.3$ clusters, e.g.,][]{farahi19}. That contributes the argument that an increase in scatter at $z<0.15$ found by \citet{Damsted23} is associated with the development of the large-scale structure. Numerical simulations demonstrate that the structure of the density field develops with time, from a smooth distribution to a very clumpy one. Statistically, this component has to be a non-linear growth component \citep{Dalal08}. Thus, our results might also be interpreted from the point of dependence of the non-linear growth of structure on the large-scale density field. This conclusion can be verified using numerical simulations. This would imply an evolution of covariance in the matter distribution between subMpc and tens of Mpc scales with redshift. Given that this term is found significant in DES analysis \citep{descl22}, it is important at $z>0.2$ as well. Our contribution is in exemplifying the underlying mechanisms, and also in eliminating projection effects as solely responsible for the observed trends. As such, searching for miracle algorithms of photometric cluster detection cannot solve the problem.

The $L_{\rm X}-\sigma_{\rm v}$ scaling relation is explored in Figs. \ref{fig: scatter LX-vdisp D cut}-\ref{fig: triangle LX-vdisp D cut} and the parameters are given in Table~A1.  \citet{Kettula15}, who studied a similar range of X-ray luminosities using weak lensing mass measurements, found a flatter slope for mergers, compared to relaxed clusters, which corresponds to a $L_{\rm X}-\sigma_{\rm v}$ slope $1.8\pm0.5$ for mergers and $2.4\pm0.3$ for relaxed systems. Flatter slopes of the relation are also obtained in studies having a small range of $\sigma_{\rm v}$ probed \citep{Gozaliasl2019}. We find a similar difference in the slopes for the extended X-ray sample, using a split based on density. However the explanation of the density-dependent effects consists in an enhanced fraction of merging clusters at lower densities, so one could state that there is a consistency in these observations.
Comparing the scaling relations based on velocity dispersion, optical luminosity exhibits the lowest scatter {(Figs. \ref{fig: scatter Lr-vdisp D cut}-\ref{fig: triangle Lr-vdisp D cut})}. However, it also reveals the largest sensitivity with respect to large-scale environment, which requires modeling, such as the one implemented in DES \citep{descl22}. We illustrate a higher variation in the optical luminosity with density via a comparison to $L_X$ in Figs. \ref{fig: scatter Lr-LX D cut}-\ref{fig: triangle Lr-LX D cut}.

\section{Discussion}\label{discussion}

We have established the sensitivity of X-ray scaling relations to the large-scale structure density. Our results can be explained from the viewpoint proposed by \cite{Manolopoulou2021}, who observed a similar change in the scaling relations, based on a lower density threshold, using voids. The idea is that the development of cool cores in clusters is linked to the absence of recent mergers, which is correlated with the density field. This offers valuable clues to a long-standing puzzle of the association of X-ray groups with the densest parts of the local web. Earlier formation time of halos in dense environments might also be at the route of the strong environmental dependence of the optical luminosity of groups.
We note that the observed trends are different from the expectation of a higher probability of interloper inclusion in a high-density environment, as this should artificially increase the measured value of velocity dispersion, while an explanation of the observed trend with respect to a decrease in the velocity dispersion is needed.

We find the optical luminosity of galaxy groups to be the best mass proxy (strictly valid only when using our definition of limiting depths and no extrapolations). It shows the lowest scatter in predicting the large-scale bias of the X-ray group sample, compared to either X-ray luminosity or velocity dispersion. A combination of optical luminosity and velocity dispersion allows for an effective cleaning of the optical group sample from outliers, with X-ray samples providing insights into the parameters of the cleaning. However, the scaling relations between the optical luminosity exhibit the highest dependency on large-scale density, which complicates clustering studies. Given that our catalogues are spectroscopic and the presence of X-ray emission confirms the groups, we believe that projection effects should be minimized. Therefore, a likely explanation of the effect is due to assembly bias \citep{wechsler02,wechsler06}, which might also be linked to galactic conformity \citep{weinmann06,kauffmann13}. We note that in dense environments, the X-ray and optical luminosity increase, so that density-dependent effects introduce a positive covariance of the scatter between optical luminosity and X-ray luminosity. They end up compensating for the negative covariance between these two properties expected from the baryonic conversion effects. 
These results point to systematic effects in using a single parameter description of the halo in combination with mass proxies. The situation can be improved by the inclusion of the halo concentration in future cosmology studies or a more complex analysis of galaxy cluster clustering, as successfully employed by DES \citep{descl22}.

\section{Conclusions}\label{conclusions}

We present catalogues of SDSS groups of galaxies with extended and compact X-ray emission and compare those to the full sample of optical groups matched in terms of the velocity dispersion.  We find large differences in the scatter of properties of groups, which we link to the influence of large-scale density on scalings relations. Groups located in high-density environments exhibit much higher optical luminosity and somewhat higher X-ray luminosity for a given velocity dispersion. The trends for the full optical sample have a larger scatter, which in part is due to the presence of low optical luminosity groups. Among the two explanations we consider (feedback or contamination), the latter better explains  the origin of the larger scatter of optical properties of the full catalogue, compared to the X-ray selected subsample. The large scatter of the X-ray luminosity, previously noted by \cite{Damsted23} has been confirmed using the extended sample, with a small reduction with redshift as well as with split on large-scale density.

We present arguments that ought to help improve our understanding of the puzzle of X-ray emission of galaxy groups at low redshifts. These include: the high importance of cool cores in X-ray detection and the dependence of the development of cool cores on the large-scale density of the cosmic web. The correlation between the optical luminosity (and therefore richness) and large-scale bias of clusters has been confirmed using the spectroscopic
membership and it cannot, therefore, be solely attributed to the projection effects. We note that our conclusions strictly apply only to the studied velocity dispersion range near 400 km/s.

\begin{acknowledgements}
S. Damsted received funding from the Finnish Emil Aaltonen Foundation. This work has been supported partially by grants PUT1627 and PRG1006 funded by the Estonian Research Council, and by the Centre of Excellence  ‘Foundations of the Universe’  (TK202). HL acknowledges financial support from the Academy of Finland, grant 311438. AF and HL thank FINCA for the travel support. AF thanks Stefano Andreon for insightful discussions. Funding for the Sloan Digital Sky Survey IV has been provided by the 
Alfred P. Sloan Foundation, the U.S. Department of Energy Office of Science, and the Participating Institutions. 

\end{acknowledgements}

   \bibliographystyle{aa} 
   \bibliography{ref} 

\begin{appendix}

\section{Redshift of $z>0.06$ sample}
\label{Appendix A}
\begin{figure}[ht]
    \centering
    \includegraphics[width=\hsize]{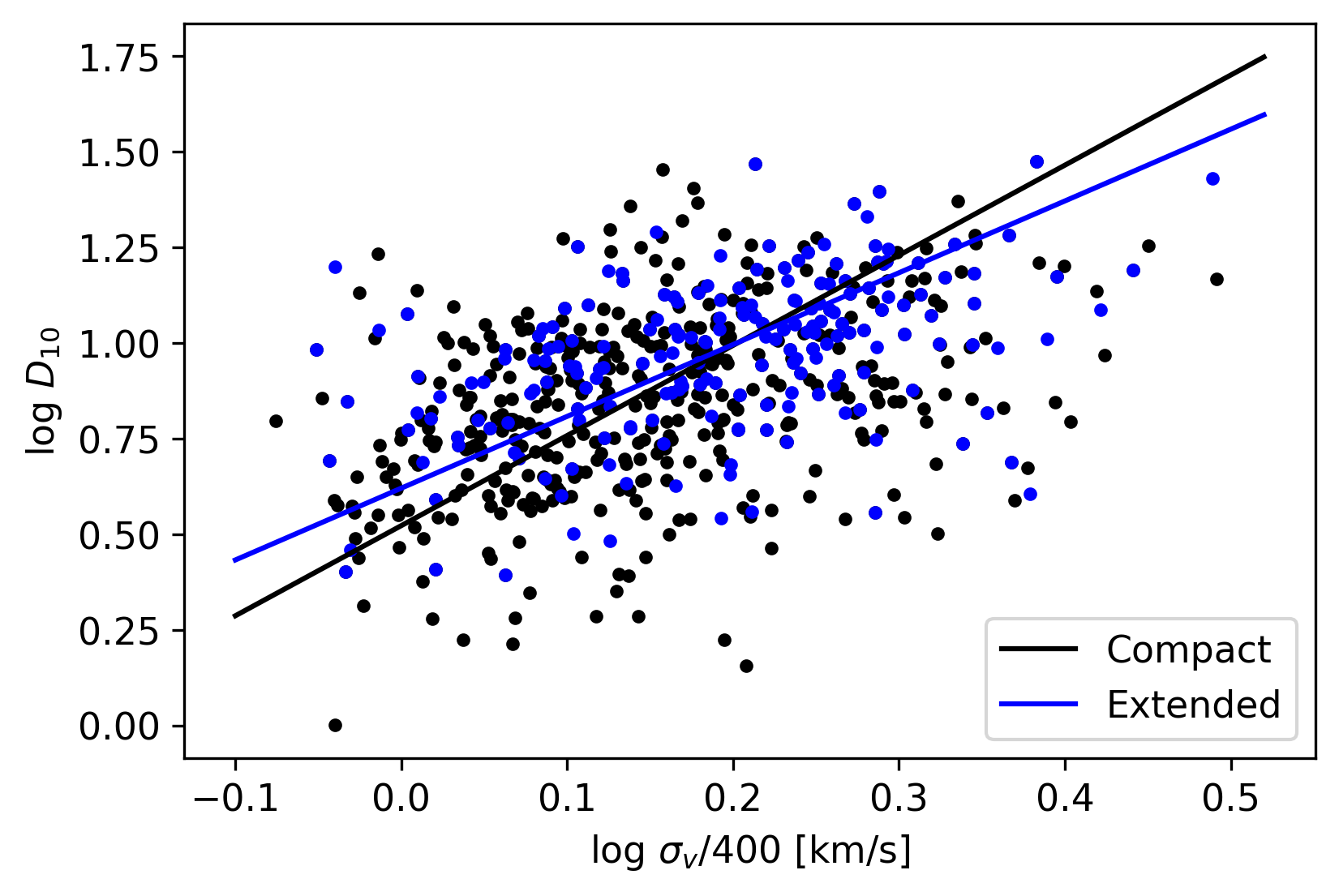}
    \caption{Large scale structure density vs. velocity dispersion. Compact and extended X-ray source samples are plotted with black and blue respectively. Similar to Fig. \ref{fig: scatter D-vdisp Lr cut}, but for $z>0.06$.}
    \label{fig: scatter D-vdisp  high z}
\end{figure}

\begin{figure}[ht]
    \centering
    \includegraphics[width=\hsize]{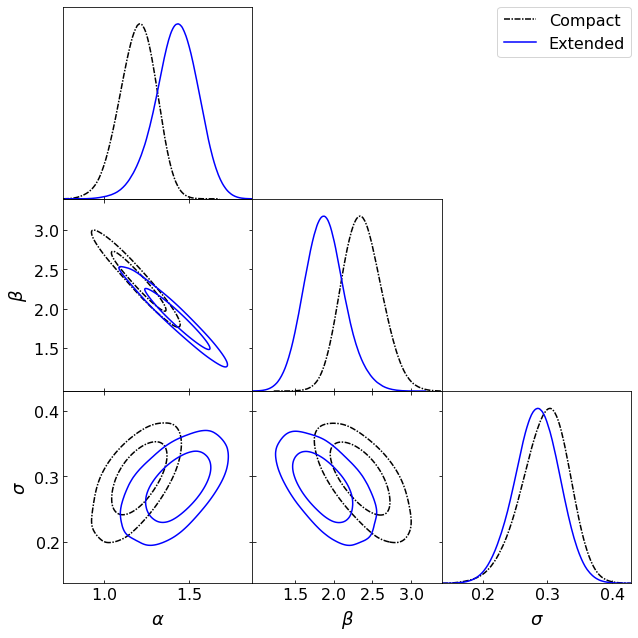}
    \caption{Large scale structure density vs. velocity dispersion. Compact and extended X-ray source samples are plotted with solid black and dash-dotted blue lines respectively. Similar to Fig. \ref{fig: triangle D-vdisp Lr cut}, but for $z>0.06$.}
    \label{fig: triangle D-vdisp  high z}
\end{figure}

\begin{figure}[ht]
    \centering
    \includegraphics[width=\hsize]{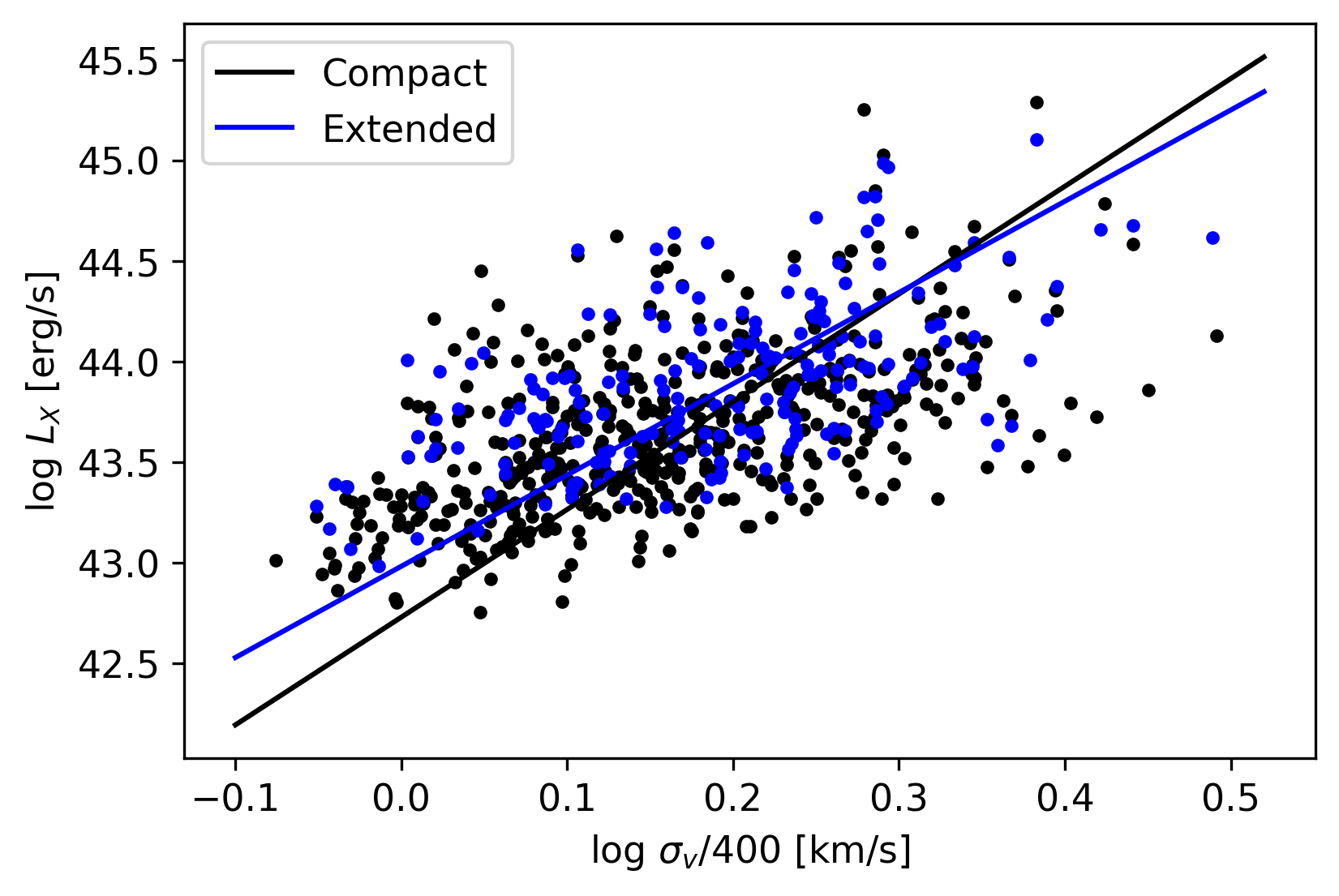}
    \caption{X-ray luminosity vs. velocity dispersion. Compact and extended X-ray source samples are plotted with black and blue respectively. Similar to Fig. \ref{fig: scatter LX-vdisp Lr cut}, but for $z>0.06$.}
    \label{fig: scatter LX-vdisp  high z}
\end{figure}

\begin{figure}[ht]
    \centering
    \includegraphics[width=\hsize]{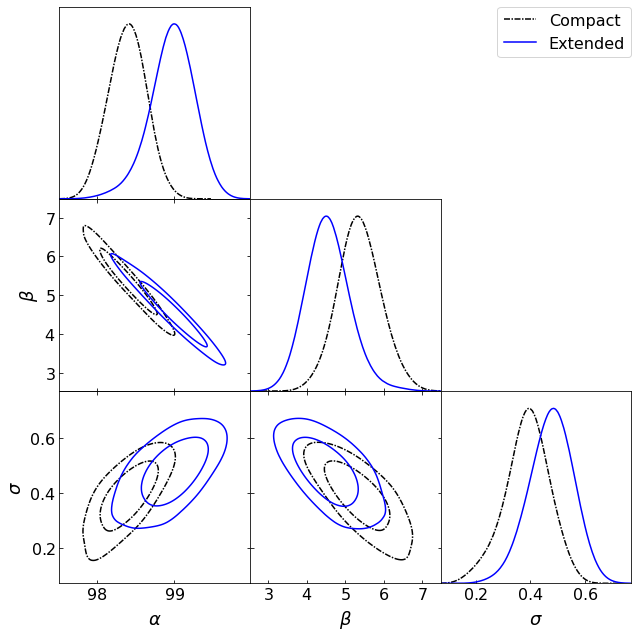}
    \caption{X-ray luminosity vs. velocity dispersion. Compact and extended X-ray source samples are plotted with solid black and dash-dotted blue lines respectively. Same as Fig. \ref{fig: triangle LX-vdisp Lr cut}, but for $z>0.06$.}
    \label{fig: triangle LX-vdisp  high z}
\end{figure}

\begin{figure}[ht]
    \centering
    \includegraphics[width=\hsize]{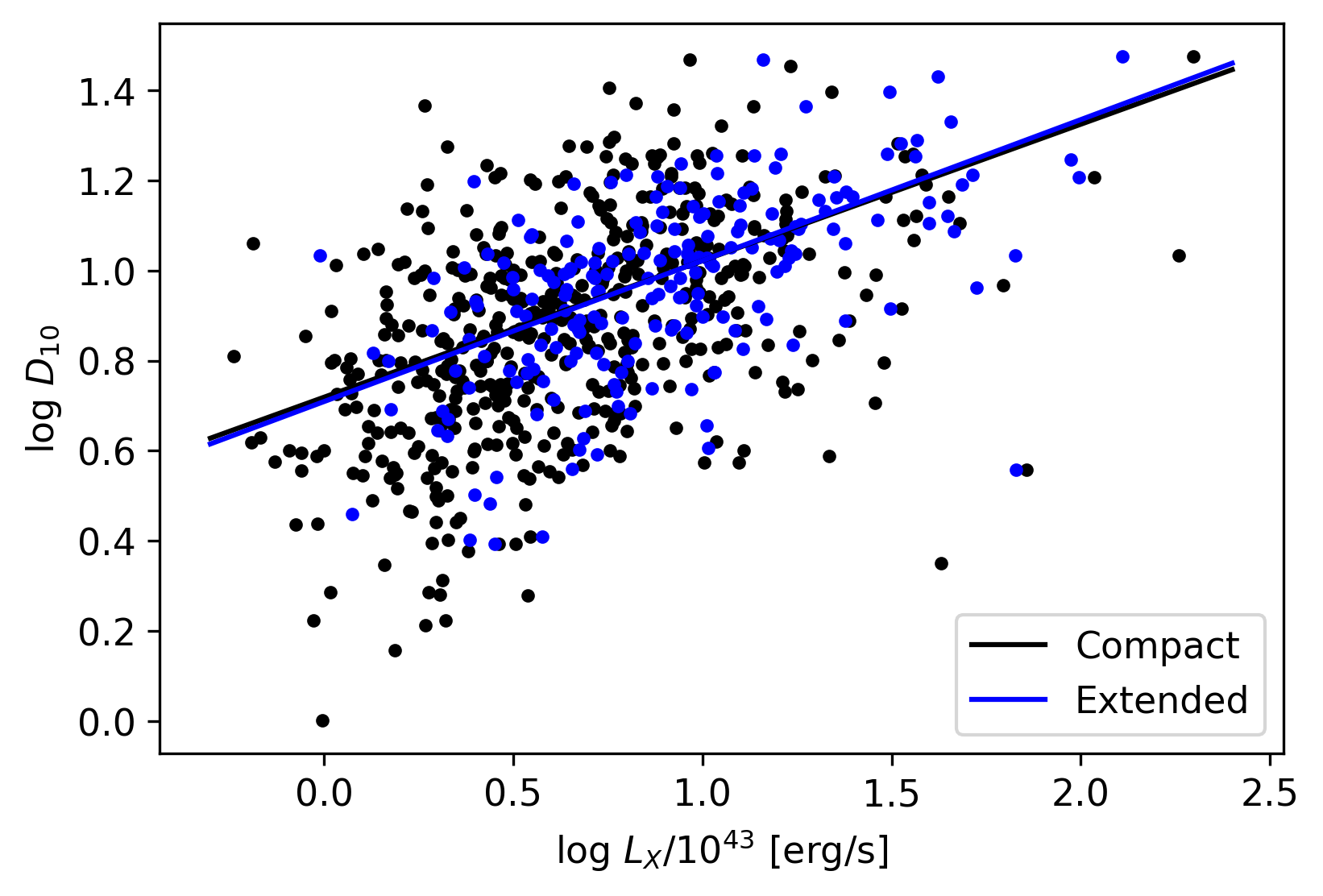}
    \caption{Large-scale structure density vs. X-ray luminosity. Same as Fig. \ref{fig: scatter D-LX Lr cut}, but for $z>0.06$.}
    \label{fig: scatter D-LX  high z}
\end{figure}

\begin{figure}[ht]
    \centering
    \includegraphics[width=\hsize]{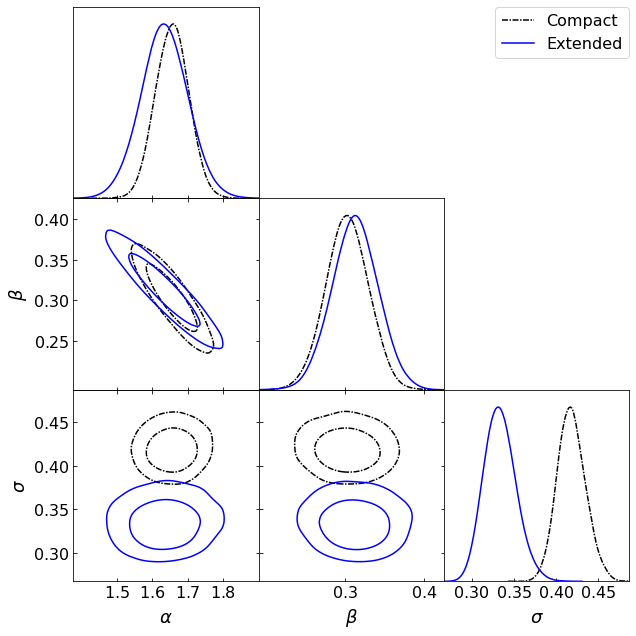}
    \caption{Large-scale structure density vs. X-ray luminosity. Same as Fig. \ref{fig: triangle D-LX Lr cut}, but for $z>0.06$.}
    \label{fig: triangle D-LX high z}
\end{figure}

\begin{figure}[ht]
    \centering
    \includegraphics[width=\hsize]{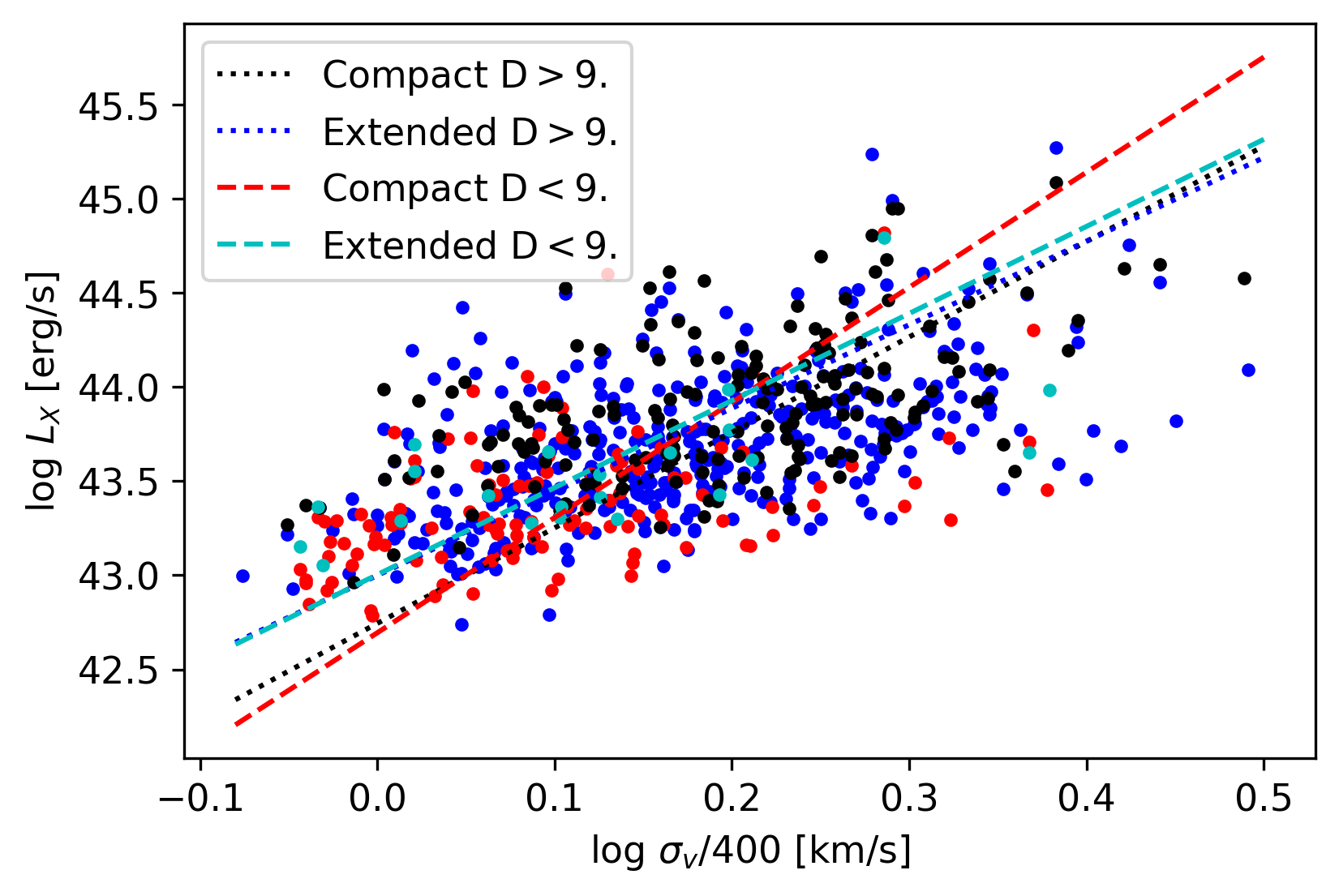}
    \caption{X-ray luminosity vs. velocity dispersion with a division in large scale structure density at $D_{10}=5.$. Same as Fig. \ref{fig: scatter LX-vdisp D cut}, but for $z>0.06$.}
    \label{fig: scatter LX-vdisp D cut high z}
\end{figure}

\begin{figure}[ht]
    \centering
    \includegraphics[width=\hsize]{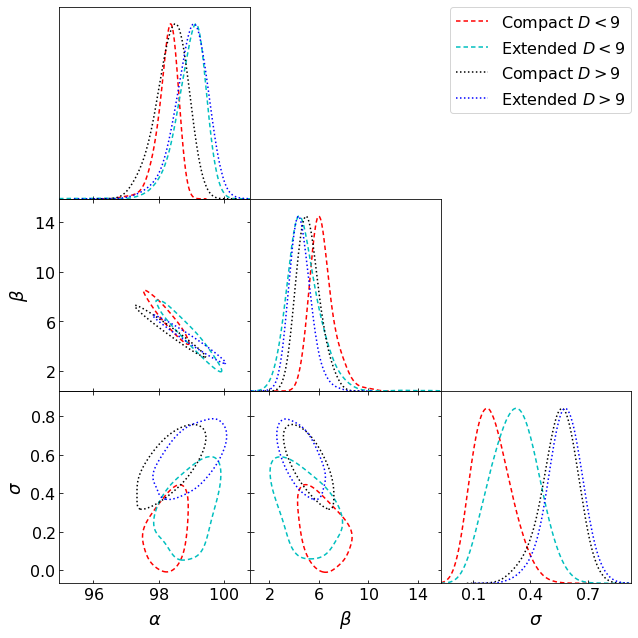}
    \caption{X-ray luminosity vs. velocity dispersion with large scale structure density cut at $D_{10}=9.$ Same as in Fig. \ref{fig: triangle LX-vdisp D cut}, but for $z>0.06$.}
    \label{fig: triangle LX-vdisp D cut high z}
\end{figure}

While we have selected $z<0.06$ for the primary analysis, here we report some of the scaling relations at $z>0.06$, using the cleaned compact and extended X-ray source samples. We applied it to both compact and extended samples the same velocity dispersion cleaning criterium given by Eq.\ref{eqn:cut2}.   
In Figs. \ref{fig: scatter D-vdisp  high z} and \ref{fig: triangle D-vdisp  high z}, we reveal that extended sources are found at marginally higher local density for a given velocity dispersion. From Figs. \ref{fig: scatter LX-vdisp  high z} and \ref{fig: triangle LX-vdisp  high z}, we conclude that extended sources have a different mass calibration, compared to the compact sources, as well as extended sources at $z<0.06$. The difference to compact sources at the same redshift might also be coming from having more low luminosity sources in the compact catalogue for a given velocity dispersion, so that the mean of the scaling relation is lower for the compact source, while the scatter is larger. 
A potential explanation can be related to the inclusion of more source flux, as at $z>0.06$ the extent of detected X-ray emission is on average twice the $R_{500c}$, which was selected as a limit of flux extrapolation. { Also, merging clusters are detected as a single extended X-ray source on those spatial scales.}
We note a reduction of scatter in the scaling relations using $L_X$ at $z>0.06$ compared to $z<0.06$ (Figs. \ref{fig: scatter LX-vdisp  high z} and \ref{fig: triangle LX-vdisp  high z} {vs. \ref{fig: scatter D-LX  high z} and} \ref{fig: triangle D-LX high z}), in agreement with the redshift-dependent trend discussed in \cite{Damsted23}. Figs. \ref{fig: scatter LX-vdisp D cut high z}-\ref{fig: triangle LX-vdisp D cut high z} demonstrate that environment-specific differences in the scaling relations are also significant for the $z>0.06$ subsamples. 

\begin{table*}
\caption{Summary of the regression analysis for the redshift z $>0.06$ sample $^a$ }             
\label{table:linmix values high z}      
\centering          
\begin{tabular}{ l c c c c c}
\hline   
Sample & Intercept $\alpha$ & Slope $\beta$ & Intrinsic scatter $\sigma$ & N clusters & Figure\\
\hline

Compact $D_{10}$ vs. $\sigma_{v}$ & 1.203 $\pm{ 0.158 }$ & 2.357 $\pm{ 0.392 }$ & 0.299 $\pm{ 0.053 }$ & 339 & \ref{fig: scatter D-vdisp  high z} \& \ref{fig: triangle D-vdisp  high z}\\

Extended $D_{10}$ vs. $\sigma_{v}$ & 1.429 $\pm{ 0.187 }$ & 1.878 $\pm{ 0.387 }$ & 0.284 $\pm{ 0.052 }$ & 175 & \ref{fig: scatter D-vdisp  high z} \& \ref{fig: triangle D-vdisp  high z}\\

Compact $L_{X}$ vs. $\sigma_{v}$ & 98.405 $\pm{ 0.368 }$ & 5.354 $\pm{ 0.852 }$ & 0.392 $\pm{ 0.121 }$ & 339 & \ref{fig: scatter LX-vdisp  high z} \& \ref{fig: triangle LX-vdisp  high z}\\

Extended $L_{X}$ vs. $\sigma_{v}$ & 98.987 $\pm{ 0.424 }$ & 4.536 $\pm{ 0.855 }$ & 0.479 $\pm{ 0.121 }$ & 175 & \ref{fig: scatter LX-vdisp  high z} \& \ref{fig: triangle LX-vdisp  high z}\\

Compact $D_{10}$ vs. $L_{X}$ & 1.656 $\pm{ 0.071 }$ & 0.303 $\pm{ 0.041 }$ & 0.418 $\pm{ 0.026 }$ & 339 & \ref{fig: scatter D-LX high z} \& \ref{fig: triangle D-LX high z}\\

Extended $D_{10}$ vs. $L_{X}$ & 1.634 $\pm{ 0.100 }$ & 0.313 $\pm{ 0.044 }$ & 0.333 $\pm{ 0.029 }$ & 175 & \ref{fig: scatter D-LX high z} \& \ref{fig: triangle D-LX high z}\\

Compact $L_{X}$ vs. $\sigma_{v}$ $D_{10}$ $< 9.$ & 98.588 $\pm{ 0.884 }$ & 5.764 $\pm{ 5.033 }$ & 0.321 $\pm{ 0.242 }$ & 192 & \ref{fig: scatter LX-vdisp D cut high z} \& \ref{fig: triangle LX-vdisp D cut high z}\\

Extended $L_{X}$ vs. $\sigma_{v}$ $D_{10}$ $< 9.$ & 99.444 $\pm{ 0.936 }$ & 2.815 $\pm{ 4.513 }$ & 0.469 $\pm{ 0.361 }$ & 70 & \ref{fig: scatter LX-vdisp D cut high z} \& \ref{fig: triangle LX-vdisp D cut high z}\\

Compact $L_{X}$ vs. $\sigma_{v}$ $D_{10}$ $> 9.$ & 98.317 $\pm{ 0.505 }$ & 5.500 $\pm{ 1.126 }$ & 0.413 $\pm{ 0.134 }$ & 147 & \ref{fig: scatter LX-vdisp D cut high z} \& \ref{fig: triangle LX-vdisp D cut high z}\\

Extended $L_{X}$ vs. $\sigma_{v}$ $D_{10}$ $> 9.$ & 98.937 $\pm{ 0.502 }$ & 4.636 $\pm{ 1.011 }$ & 0.475 $\pm{ 0.133 }$ & 105 & \ref{fig: scatter LX-vdisp D cut high z} \& \ref{fig: triangle LX-vdisp D cut high z}\\

\hline
\end{tabular}
\\
$^a$  On the X-axis: $L_X$ has been normalized by $10^{43}$ erg s$^{-1}$, $\sigma_v$ by 400 km s$^{-1}$. To account for the evolution of the scaling relations, we use $L_X E^{-1}_z$.
\end{table*}

\section{Table values}
\label{Appendix B}

{Numerical values of the analysis for the main sample with redshift $0.015<$ z $<0.06$ are presented in Table \ref{table:linmix values 1}.}

\begin{table*}
\caption{Summary of the regression analysis for the main sample with redshift $0.015<$ z $<0.06$ $^a$}             
\label{table:linmix values 1}      
\centering          
\begin{tabular}{ l c c c c c}
\hline   
Sample & Intercept $\alpha$ & Slope $\beta$ & Intrinsic scatter $\sigma$ & N groups & Figure\\
\hline

Optical $D_{10}$ vs. $L_{r}, 3\sigma$ & 1.448 $\pm{ 0.038 }$ & 0.585 $\pm{ 0.066 }$ & 0.410 $\pm{ 0.026 }$ & 279 & \ref{fig: scatter D-Lr full} \& \ref{fig: triangle D-Lr full} \\

Optical $D_{10}$ vs. Lr, full & 1.568 $\pm{ 0.043 }$ & 0.507 $\pm{ 0.088 }$ & 0.485 $\pm{ 0.032 }$ & 297 & \ref{fig: scatter D-Lr full} \& \ref{fig: triangle D-Lr full}\\
 
Compact $D_{10}$ vs. $L_{r}$ & 1.425 $\pm{ 0.064 }$ & 0.649 $\pm{ 0.098 }$ & 0.408 $\pm{ 0.042 }$ & 120 & \ref{fig: scatter D-Lr full} \& \ref{fig: triangle D-Lr full}\\

Extended $D_{10}$ vs. $L_{r}$ & 1.428 $\pm{ 0.069 }$ & 0.631 $\pm{ 0.098 }$ & 0.352 $\pm{ 0.041 }$ & 95 & \ref{fig: scatter D-Lr full} \& \ref{fig: triangle D-Lr full}\\

Optical $L_{r}$ vs. $\sigma_{v}, 3\sigma$ & 3.623 $\pm{ 0.049 }$ & 2.191 $\pm{ 0.220 }$ & 0.319 $\pm{ 0.037 }$ & 279 & \ref{fig: scatter Lr-vdisp full} \& \ref{fig: triangle Lr-vdisp full} \\

Optical $L_{r}$ vs. $\sigma_{v}$, full & 3.573 $\pm{ 0.049 }$ & 2.100 $\pm{ 0.260 }$ & 0.369 $\pm{ 0.037 }$ & 297 & \ref{fig: scatter Lr-vdisp full} \& \ref{fig: triangle Lr-vdisp full}\\

Compact $L_{r}$ vs. $\sigma_{v}$ & 3.660 $\pm{ 0.066 }$ & 1.978 $\pm{ 0.234 }$ & 0.286 $\pm{ 0.054 }$ & 120 & \ref{fig: scatter Lr-vdisp full} \& \ref{fig: triangle Lr-vdisp full} \\

Extended $L_{r}$ vs. $\sigma_{v}$ & 3.710 $\pm{ 0.071 }$ & 1.906 $\pm{ 0.236 }$ & 0.264 $\pm{ 0.054 }$ & 95 & \ref{fig: scatter Lr-vdisp full} \& \ref{fig: triangle Lr-vdisp full} \\

Optical $D_{10}$ vs. $\sigma_{v}, 3\sigma$ & 1.456 $\pm{ 0.050 }$ & 1.378 $\pm{ 0.271 }$ & 0.445 $\pm{ 0.034 }$ & 279 & \ref{fig: scatter D-vdisp Lr cut} \& \ref{fig: triangle D-vdisp Lr cut} \\

Optical $D_{10}$ vs. $\sigma_{v}$, full & 1.504 $\pm{ 0.052 }$ & 1.378 $\pm{ 0.272 }$ & 0.476 $\pm{ 0.034 }$ & 297 & \ref{fig: scatter D-vdisp Lr cut} \& \ref{fig: triangle D-vdisp Lr cut}\\

Compact $D_{10}$ vs. $\sigma_{v}$ & 1.464 $\pm{ 0.081 }$ & 1.333 $\pm{ 0.339 }$ & 0.455 $\pm{ 0.054 }$ & 120 & \ref{fig: scatter D-vdisp Lr cut} \& \ref{fig: triangle D-vdisp Lr cut}\\

Extended $D_{10}$ vs. $\sigma_{v}$  & 1.496 $\pm{ 0.086 }$ & 1.294 $\pm{ 0.319 }$ & 0.390 $\pm{ 0.053 }$ & 95 & \ref{fig: scatter D-vdisp Lr cut} \& \ref{fig: triangle D-vdisp Lr cut}\\

Compact $L_{r}$ vs. $L_{X}$ & 3.728 $\pm{ 0.072 }$ & 0.338 $\pm{ 0.074 }$ & 0.487 $\pm{ 0.050 }$ & 120 & \ref{fig: scatter Lr-LX} \& \ref{fig: triangle Lr-LX full}\\

Extended $L_{r}$ vs. $L_{X}$ & 3.796 $\pm{ 0.071 }$ & 0.366 $\pm{ 0.058 }$ & 0.400 $\pm{ 0.046 }$ & 95 & \ref{fig: scatter Lr-LX} \& \ref{fig: triangle Lr-LX full}\\

Compact $D_{10}$ vs. $L_{X}$  & 1.531 $\pm{ 0.074 }$ & 0.239 $\pm{ 0.075 }$ & 0.508 $\pm{ 0.052 }$ & 120 & \ref{fig: scatter D-LX Lr cut} \& \ref{fig: triangle D-LX Lr cut}\\

Extended $D_{10}$ vs. $L_{X}$  & 1.581 $\pm{ 0.075 }$ & 0.230 $\pm{ 0.065 }$ & 0.433 $\pm{ 0.050 }$ & 95 & \ref{fig: scatter D-LX Lr cut} \& \ref{fig: triangle D-LX Lr cut}\\

Compact $L_{X}$ vs. $\sigma_{v}$ & 98.985 $\pm{ 0.157 }$ & 3.092 $\pm{ 0.692 }$ & 0.693 $\pm{ 0.103 }$ & 120 & \ref{fig: scatter LX-vdisp Lr cut} \& \ref{fig: triangle LX-vdisp Lr cut}\\

Extended $L_{X}$ vs. $\sigma_{v}$ & 99.009 $\pm{ 0.185 }$ & 3.292 $\pm{ 0.671 }$ & 0.734 $\pm{ 0.119 }$ & 95 & \ref{fig: scatter LX-vdisp Lr cut} \& \ref{fig: triangle LX-vdisp Lr cut}\\

Compact $L_{X}$ vs. $\sigma_{v}$, D$< 5.$ & 98.588 $\pm{ 0.884 }$ & 5.764 $\pm{ 5.033 }$ & 0.321 $\pm{ 0.242 }$ & 60 & \ref{fig: scatter LX-vdisp D cut} \& \ref{fig: triangle LX-vdisp D cut}\\

Extended $L_{X}$ vs. $\sigma_{v}$, $D_{10}$ $< 5.$  & 99.444 $\pm{ 0.936 }$ & 2.815 $\pm{ 4.513 }$ & 0.469 $\pm{ 0.361 }$ & 43 & \ref{fig: scatter LX-vdisp D cut} \& \ref{fig: triangle LX-vdisp D cut}\\

Compact $L_{X}$ vs. $\sigma_{v}$, $D_{10}$ $> 5.$ & 98.317 $\pm{ 0.505 }$ & 5.500 $\pm{ 1.126 }$ & 0.413 $\pm{ 0.134 }$ & 60 & \ref{fig: scatter LX-vdisp D cut} \& \ref{fig: triangle LX-vdisp D cut}\\

Extended $L_{X}$ vs. $\sigma_{v}$, $D_{10}$ $> 5.$ & 98.937 $\pm{ 0.502 }$ & 4.636 $\pm{ 1.011 }$ & 0.475 $\pm{ 0.133 }$ & 52 & \ref{fig: scatter LX-vdisp D cut} \& \ref{fig: triangle LX-vdisp D cut}\\

Optical $L_{r}$ vs. $\sigma_{v}$, $D_{10}$ $< 5.$ & 3.506 $\pm{ 0.063 }$ & 2.130 $\pm{ 0.705 }$ & 0.276 $\pm{ 0.067 }$ & 152 & \ref{fig: scatter Lr-vdisp D cut} \& \ref{fig: triangle Lr-vdisp D cut}\\

Compact $L_{r}$ vs. $\sigma_{v}$, $D_{10}$ $< 5.$  & 3.514 $\pm{ 0.081 }$ & 1.976 $\pm{ 0.567 }$ & 0.176 $\pm{ 0.109 }$ & 60 & \ref{fig: scatter Lr-vdisp D cut} \& \ref{fig: triangle Lr-vdisp D cut}\\

Extended $L_{r}$ vs. $\sigma_{v}$, $D_{10}$ $< 5.$  & 3.554 $\pm{ 0.094 }$ & 1.788 $\pm{ 0.552 }$ & 0.209 $\pm{ 0.089 }$ & 43 & \ref{fig: scatter Lr-vdisp D cut} \& \ref{fig: triangle Lr-vdisp D cut}\\

Optical $L_{r}$ vs. $\sigma_{v}$, $D_{10}$ $> 5.$ & 3.778 $\pm{ 0.070 }$ & 1.946 $\pm{ 0.253 }$ & 0.319 $\pm{ 0.055 }$ & 127 & \ref{fig: scatter Lr-vdisp D cut} \& \ref{fig: triangle Lr-vdisp D cut}\\

Compact $L_{r}$ vs. $\sigma_{v}$, $D_{10}$ $> 5.$ & 3.849 $\pm{ 0.090 }$ & 1.708 $\pm{ 0.259 }$ & 0.274 $\pm{ 0.066 }$ & 60 & \ref{fig: scatter Lr-vdisp D cut} \& \ref{fig: triangle Lr-vdisp D cut}\\

Extended $L_{r}$ vs. $\sigma_{v}$, $D_{10}$ $> 5.$ & 3.891 $\pm{ 0.093 }$ & 1.673 $\pm{ 0.238 }$ & 0.235 $\pm{ 0.063 }$ & 52 & \ref{fig: scatter Lr-vdisp D cut} \& \ref{fig: triangle Lr-vdisp D cut}\\

Compact $L_{r}$ vs. $L_{X}$, $D_{10}$ $< 5.$  & 3.515 $\pm{ 0.081 }$ & 0.077 $\pm{ 0.123 }$ & 0.409 $\pm{ 0.060 }$ & 60 & \ref{fig: scatter Lr-LX D cut} \& \ref{fig: triangle Lr-LX D cut}\\

Extended $L_{r}$ vs. $L_{X}$, $D_{10}$ $< 5.$  & 3.624 $\pm{ 0.081 }$ & 0.265 $\pm{ 0.096 }$ & 0.343 $\pm{ 0.063 }$ & 43 & \ref{fig: scatter Lr-LX D cut} \& \ref{fig: triangle Lr-LX D cut}\\

Compact $L_{r}$ vs. $L_{X}$, $D_{10}$ $> 5.$  & 3.969 $\pm{ 0.094 }$ & 0.324 $\pm{ 0.075 }$ & 0.420 $\pm{ 0.064 }$ & 60 & \ref{fig: scatter Lr-LX D cut} \& \ref{fig: triangle Lr-LX D cut}\\

Extended $L_{r}$ vs. $L_{X}$, $D_{10}$ $> 5.$ & 3.981 $\pm{ 0.100 }$ & 0.331 $\pm{ 0.069 }$ & 0.377 $\pm{ 0.061 }$ & 52 & \ref{fig: scatter Lr-LX D cut} \& \ref{fig: triangle Lr-LX D cut}\\

\hline
\end{tabular}
\\
$^a$ . On the Y-axis, $L_r$ is in units of $10^{10} L_\odot$. On the X-axis: $L_X$ has been normalized by $10^{43}$ erg s$^{-1}$, $L_r$ by $35\times10^{10}L_\odot$ and $\sigma_v$ by 400 km s$^{-1}$. To account for the evolution of the scaling relations, we use $L_X E^{-1}_z$, $L_r E_z$.
\end{table*}

\section{Catalogs}

In this work, we have cleaned the membership of optical groups, recomputed the velocity dispersion, and obtained a uniformly defined optical luminosity. For the groups detected in X-rays we have computed the X-ray luminosity. To make the paper's results reproducible, we release all catalogues used in this work, defining the newly obtained properties of the groups.  The catalogues described in Tables C.1--4 are only available in electronic form at the CDS via anonymous ftp to cdsarc.u-strasbg.fr (130.79.128.5) or via http://cdsweb.u-strasbg.fr/cgi-bin/qcat?J/A+A/ .

\begin{table*}
    \caption{Descriptions of the columns for the clean SDSS FoF groups$^a$}
    \label{tab:catalogue_columns}
    \centering
    \begin{tabular*}{\textwidth}{l@{\extracolsep{\fill}}llr}
        \hline
        \multicolumn{1}{l}{Column} & \multicolumn{1}{l}{Unit} & \multicolumn{1}{c}{Description} & \multicolumn{1}{r}{Example} \\
        \hline

        \texttt{GROUP\_ID} (1) & & FoF group  identification number & $424$ \\
       \texttt{RA\_OPT} (2) & deg & Highest rank galaxy next X-ray peak,  right ascension (J2000) & 246.9701 \\
        \texttt{DEC\_OPT} (3) & deg & Highest rank galaxy next X-ray peak, declination (J2000) & 40.9343 \\
        \texttt{NMEM} (4) & & Number of spectroscopic members after the cleaning  & 95 \\
        \texttt{ZSPEC} (5) & & Group redshift, assigned using median value of clean members & 0.0314 \\
        \texttt{VDISP\_GAP} (6) & km s$^{-1}$ & Gapper estimate of the cluster velocity dispersion & 572.5 \\
        \texttt{EVDISP\_GAP} (7) & km s$^{-1}$ & Error on VDISP\_GAP & 53.7 \\
        \texttt{VDISP\_MAD} (8) & km s$^{-1}$ & Velocity dispersion from mean absolute deviation & 746.0\\
        \texttt{R\_200c} (9) & kpc & $R_{200c}$ radius & 1247.9181\\
        \texttt{L\_R195\_CLEAN} (10) & $L_\odot$ & L\_R\_CLEAN, using $M_{\rm r}<-19.5$ galaxy sample (z<0.06 groups)  & $93.4\times10^{10}$ \\
        \texttt{EL\_R195\_CLEAN} (11) & $L_\odot$ & Error on L\_R195\_CLEAN & 0.004$\times10^{10}$ \\
        \texttt{D\_10} (12) & $0.01227\times 10^{10}M_\odot \mbox{Mpc}^{-3}$ & Large-scale density & 16.55 \\
        \texttt{ED\_10} (13) & $0.01227\times 10^{10}M_\odot \mbox{Mpc}^{-3}$ & Error on D\_10 & 0.025 \\
        \hline
    \end{tabular*}
\\
        {$^a$ The full VAC is available online at CDS}
\end{table*}

\begin{table*}
    \caption{Descriptions of the columns for the compact X-ray catalogue of SDSS groups$^a$}
    \label{tab:catalogue_columns_com}

    \centering
    \begin{tabular*}{\textwidth}{l@{\extracolsep{\fill}}llr}
        \hline
        \multicolumn{1}{l}{Column} & \multicolumn{1}{l}{Unit} & \multicolumn{1}{c}{Description} & \multicolumn{1}{r}{Example} \\
        \hline
        \texttt{GROUP\_ID} (1) & & FoF group  identification number & $424$ \\
        \texttt{ZSPEC} (2) & & Group redshift, assigned using median value of clean members & 0.0314 \\
        \texttt{CODEX} (3) & & The CODEX cluster candidate unique identifier & 26380 \\
        \texttt{RA} (4) & deg & X-ray detection right ascension (J2000) & 246.9133 \\
        \texttt{DEC} (5) & deg & X-ray detection declination (J2000) & 40.9090 \\
        \texttt{LX0124} (6) & erg s$^{-1}$ & Luminosity in the (0.1-2.4) keV band of the cluster, aperture $R_\text{500c}$ & $1.0644\times10^{43}$ \\
        \texttt{ELX} (7) & erg s$^{-1}$ & Uncertainty on \texttt{LX0124} & $1.3159\times10^{42}$ \\
        \texttt{R200C\_DEG} (8) & deg & Apparent $R_\text{200c}$ radius of the galaxy cluster & 0.4594 \\
        \texttt{FLUX052} (9) & erg s$^{-1}$ cm$^{-2}$ & Galaxy cluster X-ray flux in the 0.5-2.0 keV band & $2.8522\times10^{-12}$ \\
        \texttt{EFLUX052} (10) & erg s$^{-1}$ cm$^{-2}$ & Uncertainty on \texttt{FLUX052} & $3.5263\times10^{-13}$ \\
        \texttt{R\_E} (11) & arcmin & Apparent radial extent of X-ray emission  & 4.29448 \\
        \hline
    \end{tabular*}
\\
        {$^a$ The full VAC is available online at CDS}
\end{table*}

\begin{table*}
    \caption{Descriptions of the columns for the extended X-ray catalogue of SDSS groups$^a$}
    \label{tab:catalogue_columns_ext}

    \centering
    \begin{tabular*}{\textwidth}{l@{\extracolsep{\fill}}llr}
        \hline
        \multicolumn{1}{l}{Column} & \multicolumn{1}{l}{Unit} & \multicolumn{1}{c}{Description} & \multicolumn{1}{r}{Example} \\
        \hline
 
        \texttt{GROUP\_ID} (1) & & FoF group  identification number & $424$ \\
        \texttt{ZSPEC} (2) & & Group redshift, assigned using median value of clean members & 0.0310 \\
        \texttt{AXES} (3) & & The AXES cluster candidate unique identifier & 93103612 \\
        \texttt{RA} (4) & deg & X-ray detection right ascension (J2000) & 247.2312 \\
        \texttt{DEC} (5) & deg & X-ray detection declination (J2000) & 40.9521 \\
        \texttt{LX0124} (6) & erg s$^{-1}$ & Luminosity in the (0.1-2.4) keV band of the cluster, aperture $R_\text{500c}$ & $1.9497\times10^{43}$ \\
        \texttt{ELX} (7) & erg s$^{-1}$ & Uncertainty on \texttt{LX0124} & $1.3893\times10^{42}$ \\
        \texttt{R200C\_DEG} (8) & deg & Apparent $R_\text{200c}$ radius of the galaxy cluster & 0.3553 \\
        \texttt{FLUX052} (9) & erg s$^{-1}$ cm$^{-2}$ & Galaxy cluster X-ray flux in the 0.5-2.0 keV band & $4.9333\times10^{-12}$ \\
        \texttt{EFLUX052} (10) & erg s$^{-1}$ cm$^{-2}$ & Uncertainty on \texttt{FLUX052} & $3.9893\times10^{-13}$ \\
        \texttt{R\_E} (11) & arcmin & Apparent radial extent of X-ray emission  & 14.9 \\
        \texttt{FLAG} (12) &  & X-ray flux contamination flag & 2 \\
        \texttt{ACOR} (13) &  & Ratio between the extrapolated and the measured flux  & 1.0 \\
        \texttt{KCOR} (14) &  & K-correction to compute the flux in the rest-frame 0.1--2.4 keV band & 1.64\\
        \hline
    \end{tabular*}
\\
        {$^a$ The full VAC is available online at CDS}
\end{table*}

The properties of optical groups are catalogued in Table~\ref{tab:catalogue_columns} with units, a short description, and an example of the data formatting.  The columns of the catalogue provide (1) FoF group ID from \citet{Tempel2017},  (2) and (3) the right ascension and declination for the epoch J2000 in degrees from \citet{Tempel2017}, (4) total number of members in the clean catalogue,  (5) the median value of cleaned cluster members, (6) the velocity dispersion calculated using the gapper estimate (GAP) in units of km s$^{-1}$, (7) the velocity dispersion computed using MAD in units of km s$^{-1}$, (8) virial radius in units of kpc, (9) $r$-band luminosity applying $M_{\rm r}<-19.5$ limit, only for $z<0.06$ groups, -99 otherwise.

The properties of compact X-ray groups are catalogued in Table~\ref{tab:catalogue_columns_com}.  The columns of the catalogue provide (1) FoF group ID from \citet{Tempel2017}, (2) redshift used in calculating X-ray properties (3) ID of compact X-ray source in the CODEX catalogue,  (4) and (5) the right ascension and declination for the epoch J2000 in degrees of the RASS X-ray detection, (6) and (7) group X-ray luminosity and associated uncertainty in units of erg s$^{-1}$ for the 0.1--2.4 keV band extrapolated to $R_\text{500c}$ from the aperture radius of $R_{\rm E}$, 
(8) and (9) the extrapolated X-ray flux in units of erg s$^{-1}$ cm$^{-2}$ for the 0.5--2.0 keV band and associated uncertainty, (10) $R_{\rm E}$: detected geometric mean radius of X-ray source ($\sqrt{ab}$, where $a$ and $b$ are axes of an ellipse describing the source), (11) source flag based on visual inspection.

The properties of extended X-ray groups are catalogued in Table~\ref{tab:catalogue_columns_ext}.  The columns of the catalogue provide: (1) FoF group ID from \citet{Tempel2017}, (2) redshift used in calculating X-ray properties (3) ID of compact X-ray source in the AXES catalogue,  (4) and (5) the right ascension and declination for the epoch J2000 in degrees of the RASS X-ray detection, (6) and (7) group X-ray luminosity and associated uncertainty in units of erg s$^{-1}$ for the 0.1--2.4 keV band extrapolated to $R_\text{500c}$ from the aperture radius of $R_{\rm E}$, (8) the $R_\text{200c}$ radius of the galaxy cluster in units of degrees, (9) and (10) the extrapolated X-ray flux in units of erg s$^{-1}$ cm$^{-2}$ for the 0.5--2.0 keV band and associated uncertainty, (11) $R_{\rm E}$: detected geometric mean radius of X-ray source ($\sqrt{ab}$, where $a$ and $b$ are axes of an ellipse describing the source), (12) contamination flag.

\begin{table*}
    \caption{Descriptions of the columns for the clean members of SDSS FoF groups$^a$}
    \label{tab:galcat}
    \centering
    \begin{tabular*}{\textwidth}{l@{\extracolsep{\fill}}llr}
        \hline
        \multicolumn{1}{l}{Column} & \multicolumn{1}{l}{Unit} & \multicolumn{1}{c}{Description} & \multicolumn{1}{r}{Example} \\
        \hline
        \texttt{GROUP\_ID} (1) & & FoF group  identification number & $3651$ \\
        \texttt{GALID } (2) & & FoF ID of the galaxy  & 213733 \\
        \texttt{SpecObjID} (3) & & SDSS ID of the galaxy  & 340080074302711808 \\
        \texttt{RA\_GAL} (4) & deg & Right ascension (J2000) & 211.8024 \\
        \texttt{DEC\_GAL} (5) & deg & Declination (J2000) & -1.012988 \\
        \texttt{ZSPEC} (6) & & Galaxy redshift & 0.0517 \\
        \texttt{L\_R} (7) & $L_\odot$ & $r$-band luminosity & $3.72243\times10^{10}$ \\
        \hline
    \end{tabular*}
\\
        {$^a$ The full VAC is available online at CDS}
\end{table*}

The format for the catalogue of the cleaned group member galaxies is described in Table~\ref{tab:galcat}. 
The columns of the catalogue provide: (1) FoF group ID from \citet{Tempel2017}, (2) FoF galaxy ID in the catalogue of \citet{Tempel2017}, (3) SDSS galaxy ID,   (4) and (5) the right ascension and declination for the epoch J2000 in degrees of galaxy,  (6) galaxy redshift with respect to CMB, (7) $r$-band luminosity of galaxy.

\end{appendix}
\end{document}